\documentclass[12pt]{article}

\usepackage[left=2.8cm,right=2.8cm,top=2.8cm,bottom=2.8cm]{geometry}

\usepackage[colorlinks=true,linkcolor=black,citecolor=black,urlcolor=blue,filecolor=black]{hyperref}

 \csname
@addtoreset\endcsname{equation}{section}

\usepackage[utf8]{inputenc}
\usepackage{amsmath}
\usepackage{amsfonts}
\usepackage{amssymb}
\usepackage{upgreek}
\usepackage{lscape}

\usepackage{soul}
\usepackage{cite}
\usepackage[usenames,table]{xcolor}
\usepackage{dsfont}
\usepackage{mathrsfs}
\newcommand{\be}{\begin{equation}}
\newcommand{\ee}{\end{equation}}

\newcommand{\cW}{{\cal W}}

\begin{document}

\vspace{30pt}

\begin{center}

%%%%%%%%%%%%%%%%%%%%%%%%%%%%%%%%%%%%%%%%%%%%%%%%%%%%%%%%

%%%%%%%%%%%%%%%%%%%%%%%%%%%%%%%%%%%%%%%%%%%%%%%%%%%%%%%%

{\Large\sc Holographic Carrollian Conformal Scalars}

\vspace{-5pt}
\par\noindent\rule{350pt}{0.4pt}

%%%%%%%%%%%%%%%%%%%%%%%%%%%%%%%%%%%%%%%%%%%%%%%%%%%%%%%%

\vspace{20pt}
{\sc 
Xavier~Bekaert${}^{\; a}$,
Andrea~Campoleoni${}^{\; b,}$\footnote{Research Associate of the Fund for Scientific Research – FNRS, Belgium.},
Simon Pekar${}^{\; c,}$\footnote{Centre National de la Recherche Scientifique, Unit\'e Mixte de Recherche UMR 7644.}
}

\vspace{8pt}
${}^a${\it\small 
Institut Denis Poisson, Unit\'e Mixte de Recherche 7013,\\ Universit\'e de Tours -- Universit\'e d'Orl\'eans -- CNRS,\\
Parc de Grandmont, 37200 Tours, France}
\vspace{4pt}

${}^b${\it\small
Service de Physique de l'Univers, Champs et Gravitation,\\
Universit{\'e} de Mons -- UMONS,\\
20 place du Parc, 7000 Mons, Belgium}
\vspace{4pt}

${}^c${\it\small
Centre de Physique Th\'eorique – CPHT\\
\'Ecole polytechnique, CNRS\\
Institut Polytechnique de Paris, 91120 Palaiseau Cedex, France}
\vspace{5pt}

{\tt\small 
\href{mailto:xavier.bekaert@lmpt.univ-tours.fr}{xavier.bekaert@lmpt.univ-tours.fr},\\
\href{mailto:andrea.campoleoni@umons.ac.be}{andrea.campoleoni@umons.ac.be},
\href{mailto:simon.pekar@polytechnique.edu}{simon.pekar@polytechnique.edu}
}

%%%%%%%%%%%%%%%%%%%%%%%%%%%%%%%%%%%%%%%%%%%%%%%%%%%%%%%%
\vspace{30pt} {\sc\large Abstract} \end{center}

\noindent

We provide holographic realisations in Minkowski spacetime of a free conformal Carrollian scalar field living at null infinity. To this end, we first show that the electric and magnetic limits of a relativistic conformal scalar are equivalent and we study the representation of the Carroll, Poincar\'e and BMS algebras that is realised on the resulting solution space. We then realise it as a quotient of the solution space of a free massless scalar in Minkowski spacetime with unusual falloff, in full analogy with the interpretation of Dirac's singleton as a shortened scalar in Anti de Sitter spacetime.

\newpage

%%%%%%%%%%%%%%%%%%%%%%%%%%%%%%%%%%%%%%%%%%%%%%%%%%%%%%%%
%%%%%%%%%%%%%%%%%%%%%%%% TEXT %%%%%%%%%%%%%%%%%%%%%%%%%%
%%%%%%%%%%%%%%%%%%%%%%%%%%%%%%%%%%%%%%%%%%%%%%%%%%%%%%%%

%\maketitle

\tableofcontents

\setcounter{footnote}{0}

%%%%%%%%%%%%%%%%%%%%%%%%%%%%%%%%%%%%
\section{Introduction}\label{sec:intro}

Field theories denoted as Carrollian have global symmetries realising the Carroll algebra, which is a contraction of the Poincar\'e algebra interpreted as a limit in which the speed of light $c$ is sent to zero \cite{Leblond, Gupta}. 
Correspondingly, Carrollian field theories are typically obtained by taking suitable $c \to 0$ limits of relativistic field theories; see, e.g., \cite{Duval:2014uoa, Bagchi:2019xfx, Gupta:2020dtl, Henneaux:2021yzg, deBoer:2021jej, Rivera-Betancour:2022lkc, Baiguera:2022lsw, Banerjee:2023jpi, deBoer:2023fnj, Koutrolikos:2023evq, Bergshoeff:2023vfd} 
(and \cite{Bergshoeff:2014jla, Chen:2021xkw, Bagchi:2022owq, Bagchi:2022eui, Figueroa-OFarrill:2023vbj} for some examples of Carrollian bottom up constructions). 
Although this limit might look unphysical, Carrollian field theories found several applications in recent years. 
Most notably, within the AdS/CFT correspondence the cosmological constant plays the role of effective speed of light in the dual boundary conformal field theory; see, e.g., \cite{Bagchi:2016bcd, Ciambelli:2018wre}. 
As a result, a limit of infinite curvature radius, $R \to \infty$, in the bulk should correspond to a Carrollian, $c \to 0$, limit on the boundary. 
This observation led to an approach to `flat holography' (i.e.\ holography for asymptotically flat spacetimes) aiming at describing the gravitational dynamics by means of (conformal) Carrollian field theories defined at null infinity; see, e.g., \cite{Dappiaggi:2005ci, Bagchi:2016bcd, Ciambelli:2018wre, Donnay:2022aba, Bagchi:2022emh, Donnay:2022wvx, Salzer:2023jqv, Nguyen:2023vfz, Campoleoni:2023fug, Bagchi:2023cen, Mason:2023mti}. 
It has also been realised that the Carrollian path towards flat space holography actually provides an equivalent, but complementary, formulation of the program going under the name of `celestial holography' (see, e.g., \cite{Pasterski:2021raf, Donnay:2023mrd} for recent reviews).

In accordance with the previous picture, we denote as conformal Carrollian field theories the $c \to 0$ limit of relativistic field theories whose global symmetries contain the conformal algebra. 
In this limit the global conformal algebra contracts to an extension of the Carroll algebra that is isomorphic to the Poincar\'e algebra in one more dimension, while the corresponding field theories typically display enhanced global symmetries realising the Bondi-Metzner-Sachs (BMS) algebra (or an extension thereof). 
The latter also describes the asymptotic symmetries of gravity in asymptotically flat spacetimes of one more dimension, thus further supporting the `Carrollian holography' proposal. Motivated by the role of conformal Carrollian field theories as putative holographic duals of gravitational theories on asymptotically flat spacetimes, in this work we study their simplest instance, given by the $c \to 0$ limit of a relativistic conformal scalar on the Lorentzian manifold $\mathbb R \times S^d$, interpreted as the conformal boundary of a $(d+2)-$dimensional Anti de Sitter (AdS) space. 
This is the most manageable Carrollian field theory, so that analysing it in detail is expected to shed light on the general, peculiar, properties of (conformal) Carrollian field theories; see also \cite{deBoer:2021jej, deBoer:2023fnj} for previous works in this direction.

Aside from providing a handy example of a Carrollian field theory, this model also has the virtue of admitting a potential holographic interpretation. 
A relativistic free scalar indeed exhibits an infinite number of Noether currents besides those associated to conformal symmetry. In the usual AdS/CFT setup, these are expected to couple to the boundary values of gauge fields of arbitrary spin \cite{Sundborg:2000wp, Witten_talk, Mikhailov:2002bp, Sezgin:2002rt, Klebanov:2002ja}. Concrete realisations of this holographic scenario have been developed over the years (see, e.g., \cite{Giombi:2016ejx} for a review), so that the $c \to 0$ limit of a conformally coupled scalar is expected to enter a holographic description of higher-spin theories in flat space, if any.
The latter scenario has been long overlooked because, in the common lore, higher-spin gravity theories are considered to be inconsistent in Minkowski spacetime due to a series of no-go theorems severely constraining their interactions; see, e.g., \cite{Bekaert:2010hw} for a review. 
In spite of these constraints, some interacting higher-spin theories have been however defined on flat manifolds with Euclidean or split signature \cite{Ponomarev:2016lrm, Krasnov:2021nsq} and various steps towards their holographic description have been performed in \cite{Ren:2022sws, Ponomarev:2022ryp, Ponomarev:2022qkx, Monteiro:2022xwq}. 
Recent progress also suggests that a counterpart of Vasiliev's equations, describing higher-spin interactions in AdS, could be defined in Minkowski spacetime too. This proposal relies on the observation that the infinite-dimensional Lie algebra underlying Vasiliev's equations admits a contraction whose linearised curvatures can be used to describe the free dynamics of fields of arbitrary fields \cite{Campoleoni:2021blr, Boulanger:2023prx}. 
The latter step mimics Cartan's approach to Einstein's gravity and it is one of the key ingredients in Vasiliev's construction.
For the scope of this paper, the main manifestation of this scenario is the fact that the Carrollian limit of a relativistic conformal scalar is invariant under a huge  algebra of global symmetries that contains as a subalgebra the previous higher-spin algebra on Minkowski spacetime \cite{Bekaert:2022oeh}.

Our analysis of the ultra-relativistic limit of a scalar conformally coupled to the background $\mathbb R \times S^d$ focuses on various aspects. 
First of all, the Carrollian limit of a relativistic field theory can usually be taken in two ways that were dubbed `electric' and `magnetic' in \cite{Henneaux:2021yzg}, in analogy with the electric- and magnetic-field dominated limits of Maxwell's theory \cite{Duval:2014uoa}. The same applies to the conformal scalar field, and its electric (or time-like) and magnetic (or space-like) limits were defined in \cite{Henneaux:2021yzg, Rivera-Betancour:2022lkc, Baiguera:2022lsw, Bekaert:2022oeh}. 
On the other hand, in Section~\ref{electric/magnetic} we point out that in the case where no mass parameter is present, the two limits are actually equivalent both on-shell and off-shell (up to the non-local inversion of a shifted Laplacian that we will discuss in detail). 
The equivalence between electric and magnetic limits is also revisited in Appendix~\ref{sec:duality} from the viewpoint of electric-magnetic duality.

In Section~\ref{groupth}, we then study the properties of the representation of the Carroll algebra that is realised on the space of solutions of the conformal Carrollian scalar field theory. 
This representation lifts to a non-unitary indecomposable representation of the Poincar\'e and BMS algebras in one more dimension that we call \emph{simpleton}. 
Besides the fact that the corresponding Carrollian field theory may qualify as the simplest possible one, this name is motivated by its close analogies with Dirac's `remarkable representation' \cite{Dirac:1963ta} called singleton, that also plays a crucial role in higher-spin holography. 
After reviewing the definition of Dirac's singleton, we highlight these similarities in Section~\ref{sec:limit}.
In particular, we stress that (super)translation generators are realised as nilpotent operators on the simpleton representation and that this property, $\hat{P}_a\hat{P}_b\, |\text{simpleton} \rangle = 0$, is inherited from the similar property, $(\hat{P}_{(a}\hat{P}_{b)} + \hat{J}_{c(a}\hat{J}_{b)}{}^c/R^2 )|\text{singleton}\rangle = 0$, that holds for the singleton. 
In general, we show that the simpleton can be considered as a flat/Carrollian limit of the singleton.

The option to interpret the limit of the singleton leading to the simpleton as a Carrollian limit exploits the realisation of the former as the space of solutions of a relativistic conformal scalar in $d+1$ dimensions.
The option to interpret the same limit as a flat limit exploits instead the `holographic' realisation \cite{Fronsdal:1978vb, Angelopoulos:1980wg} of Dirac's singleton as a shortened scalar in AdS$_{d+2}$. 
By this we mean that the singleton representation can be realised as a quotient of the space of solutions of a Klein-Gordon equation with a fine-tuned mass by an infinite-dimensional invariant subspace. 
In Section~\ref{sec:Minkowski}, we show that a similar description in terms of a quotient of the space of solutions of the d'Alembert equation in Minkowski spacetime applies to the simpleton too. 
This provides a concrete holographic description of an on-shell conformal Carrollian scalar field in Minkowski spacetime, that can be obtained from a flat limit of the corresponding holographic description of the singleton in AdS.\footnote{The effect of a limit of vanishing cosmological constant on the singleton representation was already discussed in \cite{Flato:1978qz}, where it was however concluded that it should give the zero-momentum representation of the Poincar\'e algebra. 
As we will see, the latter however only corresponds to an invariant subspace of our simpleton representation in which (super)translations are nilpotent, but do not act trivially (see also \cite{Ponomarev:2021xdq}).} 
We also show how the same solution space can be described in terms of a doublet of homogeneous fields in Minkowski spacetime, which makes manifest the nilpotency of (super)translations.

The $c \to 0$ limit not only affects the field theory, but also turns the Lorentzian structure of the background manifold into a degenerate metric structure \cite{Henneaux:1979vn}, nowadays called Carrollian structure \cite{Duval:2014uoa}. 
The Carrollian manifold on which our scalar theory is defined can be interpreted either as the (future or past) null infinity of Minkowski spacetime or as the boundary of a Carrollian manifold of one more dimension called AdS-Carroll spacetime. 
Involving a Carrollian limit on both the bulk and the boundary, the latter picture admits simple holographic descriptions that we discuss in Section~\ref{sec:ambient} by taking advantage of ambient space techniques. Last but not least, in Appendix~\ref{sec:symmetries-magnetic} we study all global symmetries of the magnetic action for a conformal Carrollian scalar, including symmetry generators of order higher than one (aka `higher-spin' symmetries). 
This complements our previous study of the symmetries of the electric action \cite{Bekaert:2022oeh} and, consistently with the off-shell equivalence of the two theories, we show explicitly that the global symmetries of the two actions agree.

%%%%%%%%%%%%%%%%%%%%%%%%%%%%%%%%%%%%
\section{Boundary definitions} \label{sec:boundary}

We start by reviewing two formulations, dubbed electric and magnetic \cite{Henneaux:2021yzg, deBoer:2021jej}, of a conformal Carrollian scalar field defined on the manifold $\mathscr{I}_{d+1} \cong \mathbb R \times S^d\,$. 
As we will review, the latter manifold can be seen either as the (past or future) null infinity of Minkowski spacetime $\mathbb{R}^{d+1,1}\,$, or as the ultra-relativistic limit of spatial infinity of Anti de Sitter spacetime AdS$_{d+2}\,$. 
We then point out that the electric and magnetic formulations admit the same solution space and that they are in fact equivalent, even off-shell, up to a non-local inversion of a shifted Laplacian on the celestial sphere.
We conclude this section by discussing the properties of the representation of the Carroll, Poincar\'e and BMS algebras realised on the solution space of a conformal Carrollian scalar.

%%%%%%%%%%%%%%%%%%
\subsection{Electric and magnetic descriptions}
\label{electric/magnetic}

The electric and magnetic formulations of a conformal Carrollian scalar are both obtained by taking suitable limits of the action for a free relativistic complex conformal scalar on the Lorentzian manifold $\mathbb R \times S^d$, which is the conformal boundary of the universal covering of AdS$_{d+2}$. 
The latter reads
\begin{equation} \label{relativisticS}
S_\text{rel}[\phi] = \frac12 \int {\rm d}u\,{\rm d}^d \mathbf x\ \sqrt{\gamma}\, \left(c^{-2}\,|\dot\phi|^2 +\phi^*\hat\nabla^2 \phi\right) ,
\end{equation}
where $\mathbf x$ denote $d$ angular coordinates on the celestial sphere, $\gamma$ is the round-sphere metric on it and a dot stands for $\partial_u\,$. 
Moreover, $c$ is the velocity of light and
\begin{equation}
\label{shiftedLaplacian}
    \hat\nabla^2 := \nabla_{S^d}^2 - \left(\frac{d-1}{2}\right)^2
\end{equation}
with $\nabla_{S^d}^2$ the covariant Laplacian on the sphere $S^d$. 
This operator arises in the conformal completion of the d'Alembertian on $\mathbb R \times S^d$:
\be
\hat{\Box}_{{}_{\mathbb R \times S^d}}:=\Box_{{}_{\mathbb R \times S^d}}-\left(\frac{d-1}2\right)^2
=-c^{-2}\partial_u{}^2+\hat{\nabla}^2 \, .
\ee
For the ensuing discussion, it is convenient to also rewrite the action \eqref{relativisticS} in a first-order form by introducing the auxiliary field $\pi$, which corresponds to the canonical momentum up to a $\sqrt{\gamma}$ factor,
\begin{equation} \label{first-order-S}
    S_\text{rel}[\phi,\pi] = \frac12 \int {\rm d}u\,{\rm d}^d \mathbf x\ \sqrt{\gamma} \left(\pi^*\dot\phi + \pi\dot\phi^* + \phi^* \hat\nabla^2 \phi - c^2 |\pi|^2 \right) .
\end{equation}

The electric action for a (complex) conformal Carrollian scalar field $\varphi$ with scaling dimension $\Delta_\varphi = \frac{d-1}{2}$, living on the Carrollian manifold $\mathscr{I}_{d+1}$, is the ultra-relativistic ($c \to 0$) limit of the second-order relativistic action \eqref{relativisticS}, taken after having performed the rescaling $\phi=c\,\varphi$. 
It reads 
\be \label{eq:boundary-electric}
    S_\text{el}[\varphi] = \frac12 \int {\rm d}u\,{\rm d}^d \mathbf x\ \sqrt{\gamma}\, |\dot\varphi|^2 \,.
\ee
This action is invariant under Carrollian transformations and their conformal extension (which realise the Poincar\'e algebra in one more dimension, as reviewed in Appendix~\ref{sec:limits-bulk-bdy}). 
In fact, it is invariant under an infinite-dimensional extension of the conformal-Carroll/Poincar\'e isometries containing the BMS algebra \cite{Bagchi:2019xfx}, including the super-rotations of \cite{Campiglia:2014yka}, together with additional transformations described in \cite{Bekaert:2022oeh} and reviewed in Appendix~\ref{sec:symmetries-magnetic}.

The Euler-Lagrange equation for the electric action \eqref{eq:boundary-electric} is
\be \label{eq:boundary-electric-eom}
    \ddot \varphi = 0
\ee
and its general solution reads
\be\label{gensol}
\varphi(u,\mathbf x)=\varphi_-(\mathbf x)+u\,\varphi_+(\mathbf x)\,,
\ee
where $\varphi_\pm(\mathbf x)$ are conformal primary fields on the celestial sphere $S^d$ with scaling dimensions $\Delta_\pm = \frac{d\pm1}{2}\,$. 
These scalar conformal primary fields with all their descendants span an irreducible module (i.e.\ a representation space), often denoted $\mathcal{D}(\Delta_\pm,0)$, of the algebra $\mathfrak{so}(d+1,1)$ of global conformal isometries of the celestial sphere $S^d$. 
We will denote the representation of the Carroll, Poincar\'e and BMS algebras realised on the solution space~\eqref{gensol} as \emph{simpleton}.

The decomposition \eqref{gensol} makes manifest the triangular structure of the action of super-translations $u\to u+f(\mathbf x)$ (where $f$ has scaling dimension $-1$) on the space of solutions $\varphi(u,\mathbf x)$: on the one hand, the field $\varphi_+(\mathbf x)$ is inert under super-translations while, on the other hand, the field $\varphi_-(\mathbf x)$ receives a contribution from $\varphi_+(\mathbf x)$. 
More precisely, this doublet transforms as $\varphi_-(\mathbf x)\to \varphi_-(\mathbf x)+f(\mathbf x)\,\varphi_+(\mathbf x)$ and $\varphi_+(\mathbf x)\to \varphi_+(\mathbf x)$.
Grouping the conformal primary fields $\varphi_\pm(\mathbf x)$ in a doublet
\be \label{eq:electric-doublet}
\boldsymbol{\varphi}(\mathbf x) = \begin{pmatrix} \varphi_-(\mathbf x) \\ \varphi_+(\mathbf x) \end{pmatrix} ,
\ee
finite super-translations have a manifestly upper triangular structure
\be\label{uppertriangular}
\boldsymbol{\varphi}(\mathbf x) \to \begin{pmatrix} 1 & f(\mathbf x)\\ 0 & 1 \end{pmatrix}
\boldsymbol{\varphi}(\mathbf x) \,,
\ee
so that, infinitesimally, they have a strictly upper triangular form. 
In particular, it is clear from \eqref{gensol} that the Hamiltonian generator, realised as $\hat{P}_u=i\partial_u$ on the scalar $\varphi(u,\mathbf x)$, is realised as a strictly upper triangular $2\times 2$ matrix on the doublet $\boldsymbol{\varphi}(\mathbf x)$. Accordingly, infinitesimal (super)translations are realised on the simpleton as nilpotent operators (see Section~\ref{groupth} for more comments on this point).

The magnetic action, on the other hand, can be obtained by taking the $c \to 0$ limit of the first-order relativistic action \eqref{first-order-S} and reads 
\begin{equation} \label{eq:boundary-magnetic}
    S_\text{mag}[\phi,\pi] = \frac12 \int {\rm d}u\,{\rm d}^d \mathbf x\, \sqrt{\gamma} \left( \pi^* \dot \phi + \pi \,\dot \phi^* + \phi^* \hat\nabla^2 \phi \right) ,
\end{equation}
where the fields $\phi$ and $\pi$ have scaling dimensions $\Delta_\phi = \Delta_- = \frac{d-1}{2}$ and $\Delta_\pi = \Delta_+ = \frac{d+1}{2}$.
Notice that the shifted Laplacian $\hat\nabla^2$ introduced in \eqref{shiftedLaplacian} also enters this action and differs from the Laplace-Yamabe operator (aka conformal Laplacian) on $S^d$, as was already noticed in \cite{Rivera-Betancour:2022lkc}. 
This is because both the relativistic and Carrollian theories are formulated on the manifold $\mathscr{I}_{d+1}$ (with the appropriate metric structures) and not on $S^d$.\footnote{In fact, the shifted Laplacian \eqref{shiftedLaplacian} would also arise from the Galilean limit of the conformal completion of the d'Alembertian on $\mathbb R \times S^d$.}
In the following, this will play a key role in understanding the holographic nature of the magnetic theory. 
The expression for the shifted Laplacian $\hat\nabla^2$ can also be derived by imposing that the action \eqref{eq:boundary-magnetic} be invariant under Weyl rescalings of the Carrollian geometry, see \cite{Baiguera:2022lsw} where it was also checked that \eqref{eq:boundary-magnetic} is invariant under conformal Carroll boosts too. 
Furthermore, in Appendix~\ref{sec:symmetries-magnetic} we will show that the magnetic action display the same additional infinite-dimensional symmetries as the electric one.

The equations of motion for the magnetic action \eqref{eq:boundary-magnetic} are given by
\begin{equation} \label{eq:boundary-magnetic-eom}
    \dot\phi = 0 \,,\quad \dot \pi = \hat\nabla^2 \phi \,.
\end{equation}
Their general solution is
\be \label{eq:magnetic-on-shell}
\phi(u,\mathbf x)=\psi_-(\mathbf x)\,,\qquad \pi(u,\mathbf x)=\psi_+(\mathbf x)+u\,\hat\nabla^2\psi_-(\mathbf x)\,,
\ee
where $\psi_\pm(\mathbf x)$ are conformal primary fields on $S^d$ with the same scaling dimension $\Delta_\pm = \frac{d\pm1}{2}$ as in \eqref{gensol}. 
This makes manifest that the two theories, electric and magnetic, share the same space of solutions although, 
grouping the pair of conformal primary fields $\psi_\pm(\mathbf x)$ in a doublet
\be \label{eq:magnetic-doublet}
\boldsymbol{\psi}(\mathbf x) = \begin{pmatrix} \psi_-(\mathbf x) \\ \psi_+(\mathbf x) \end{pmatrix} ,
\ee
one can note that the action of super-translations in the magnetic formulation has a lower triangular structure
\be\label{lowertriangular}
\boldsymbol{\psi}(\mathbf x)\, \to\, \begin{pmatrix} 1 & &0&\\ f(\mathbf x) \hat\nabla^2 + \partial^i f(\mathbf x)\, \nabla_{\!i} & &1& \end{pmatrix}
\boldsymbol{\psi}(\mathbf x)
\ee
which looks very different from the transformation \eqref{uppertriangular} in the electric description (see also Appendix~\ref{sec:symmetries-magnetic}).

In fact, in spite of this apparent discrepancy,
the classical `electric' and `magnetic' theories on $\mathscr{I}_{d+1}$ are actually equivalent, both on-shell and off-shell. 
More precisely, they are the same up to a non-local inversion of the shifted Laplacian $\hat \nabla^2$. 
This equivalence is perhaps to be expected, since the electric and magnetic limits of electromagnetism are, in spacetime dimension four, related by an electric-magnetic duality transformation. 
This is particularly manifest on-shell \cite{Duval:2014uoa}, but it also holds off-shell. 
More on this issue and its generalisation to any spin and even dimension in Appendix~\ref{sec:duality}.

For the simpleton on $\mathscr{I}_{d+1}$, the equivalence holds in any\footnote{The equivalence also holds in $d=1$ except for the zero-mode $\hat\nabla^2 \phi = \partial_\theta{}^2 \phi = 0$.} dimension $d \geqslant 2$ and the explicit relation between the electric and magnetic formulations is as follows. 
By eliminating $\phi$ through its own equation of motion, we get the non-local relation
\be \label{eq:non-local}
\phi = \frac{1}{\,\hat\nabla^2}\,\dot\pi \,,
\ee
which shows that $\phi$ is an auxiliary field. 
Plugging it back in the magnetic action, we arrive at the electric action $S_\text{el}[\varphi]$ after defining
\be \label{eq:electric-from-magnetic}
\varphi = \frac{1}{\sqrt{-\hat\nabla^2}} \,\pi 
\ee
and ignoring boundary terms.
This inversion of the shifted Laplacian on the round sphere metric is always possible when $d \geqslant 2$, as can be seen from the action of $\hat\nabla^2$ on a spherical harmonic $Y_{\ell \geqslant 0}$ of the $d$-dimensional sphere:
\begin{equation}
    -\hat\nabla^2 Y_\ell = \ell(\ell + d-1) Y_\ell + \left( \frac{d-1}{2}\right)^2 Y_\ell = \left(\ell + \frac{d-1}{2}\right)^2 Y_\ell \,,
\end{equation}
where $\ell + \frac{d-1}{2} > 0$ as soon as $d > 1$. 
The general solutions in the electric and magnetic formulations are matched by applying \eqref{eq:electric-from-magnetic} to the second relation in \eqref{eq:magnetic-on-shell}:
\be\label{vargensol}
\varphi(u,\mathbf x)=\frac{1}{\sqrt{-\hat\nabla^2}}\,\psi_+(\mathbf x)-u\,\sqrt{-\hat\nabla^2}\,\psi_-(\mathbf x)
=\varphi_-(\mathbf x)+u\,\varphi_+(\mathbf x)\,,
\ee
where we introduced $\varphi_\mp\left(\mathbf x\right)=\pm\,(-\hat\nabla^2)^{\mp\frac12}\psi_\pm(\mathbf x)$ in order to fit the form of the general solution \eqref{gensol} in the electric formulation. 
Note that $\varphi_-$ is related to $\psi_+$ and $\varphi_+$ to $\psi_-$, thus explaining why the action of (super)translations is upper-triangular in one case and lower-triangular in the other.

%%%%%%%%%%%%%%%%%%
\subsection{Group-theoretical description}\label{groupth}

The Carroll algebra (spanned by the isometries of flat Carroll spacetime) arises as the ultra-relativistic ($c \to 0$) contraction of the Poincar\'e algebra: $\mathfrak{iso}(d,1)\stackrel{c\to0}{\longrightarrow}\mathfrak{carr}(d,1)$. 
Applying the same ultra-relativistic contraction to the global conformal algebra, one gets a subalgebra of the conformal isometries of flat Carroll spacetime which is isomorphic to a Poincar\'e algebra in one dimension higher: $\mathfrak{so}(d+1,2)\stackrel{c\to0}{\longrightarrow}\mathfrak{iso}(d+1,1)$. 
The whole algebra of conformal isometries of flat Carroll spacetime is then isomorphic to the full BMS algebra in one more dimension and this is also true for $\mathscr{I}_{d+1} \cong \mathbb R \times S^d$ \cite{Duval:2014uva}.
This algebraic pattern is at the origin of the `Carrollian holography' proposal mentioned in the introduction:
within the AdS/CFT correspondence, the previous isomorphisms signal that a flat limit ($R\to\infty$) in the bulk of AdS spacetime is equivalent to a Carrollian limit ($c \to 0$) on the boundary; see, e.g., \cite{Ciambelli:2018wre}. 
We review the isomorphism between the Carrollian conformal algebra and the Poincar\'e algebra in Appendix~\ref{sec:limits-bulk-bdy}. 
In Section~\ref{sec:limit-field}, a holographic description of a scalar field in AdS will also make manifest that the inverse AdS curvature radius plays the role of effective speed of light on the conformal boundary.

In this section, motivated by the previous isomorphisms, we describe the vector space of solutions of the linear equation \eqref{eq:boundary-electric-eom} or, equivalently, of the linear system \eqref{eq:boundary-magnetic-eom} as a module of the various relevant algebras (Carroll, Lorentz, Poincar\'e and BMS).\footnote{We focus on algebras for the sake of simplicity, to avoid topological subtleties.} 
As anticipated, we will refer to it as the (on-shell) `simpleton', without distinguishing between the electric and magnetic formulations since they lead to identical representations. 

In accordance with the previous discussion, the simpleton carries a representation of the algebra of isometries of the Carrollian manifold $\mathscr{I}_{d+1}\cong \mathbb{R}\times S^d$. 
This representation lifts to a representation of the global conformal Carroll algebra, aka Poincar\'e algebra $\mathfrak{iso}(d+1,1)$, and further to the BMS algebra $\mathfrak{bms}_{d+2}$ (and even further to the generalised BMS and Newman-Unti algebras, cf.~Appendix~\ref{sec:symmetries-magnetic}).
Note that the representation carried by the simpleton is neither unitary nor irreducible. 
In fact, this representation is indecomposable (i.e.\ reducible but not fully reducible) and this implies that the representation is \textit{not} unitarisable.\footnote{Remember that all unitary representations are fully reducible (by taking the orthogonal complement of the invariant subspace). Thus, by contraposition, an indecomposable representation cannot be unitary.} 
Let us discuss in more details the indecomposable structure of this representation.

The simpleton representation is \textit{reducible} because it possesses an invariant subspace: the space spanned by solutions of the equation $\dot{\varphi}=0$ is clearly a subspace of the electric equation of motion \eqref{eq:boundary-electric-eom}. 
In the corresponding magnetic formulation, this subspace is to be understood as the subspace of solutions of the system \eqref{eq:boundary-magnetic-eom} that satisfy  
$\phi=0$ (equivalently, $\dot{\pi}=0$). 
This invariant subspace carries the \textit{zero-energy representation} of the Carroll algebra $\mathfrak{carr}(d,1)$  since the Carroll Hamiltonian is realised as $\hat{H}=i\,\partial_u$. 
This zero-energy representation is \textit{not} faithful, since time translations are realised trivially (on-shell). However, it is unitary and irreducible, see e.g. \cite{Figueroa-OFarrill:2023vbj}. 
It lifts to an (unfaithful) unitary irreducible representation of the Poincar\'e algebra \mbox{$\mathfrak{iso}(d+1,1)$} or of the BMS algebra $\mathfrak{bms}_{d+2}$, where (super)translations act trivially. 
Accordingly, it is sometimes called the zero-(super)momentum representation in the classification of the unitary irreducible representations of the Poincar\'e \cite{Wigner:1939cj} (or BMS \cite{McCarthy01,McCarthy317}) algebra.

The simpleton representation is also \textit{indecomposable} because the Carroll Hamiltonian \mbox{$\hat{H}=i\,\partial_u$} and, more generally, the generators of (super)translations $f(\mathbf{x})\partial_u$ have a strictly triangular structure. 
In particular, the (super)translation generators act on the simpleton as nilpotent operators (which is obvious in the electric formulation, since $\ddot{\varphi}=0$). 
The strictly triangular structure of infinitesimal super-translations is manifest in terms of the doublet \eqref{eq:electric-doublet} of conformal primary fields $\varphi_\pm$. The fact that the  (super)translation generators are nilpotent shows that the simpleton does not admit a (super)momentum basis of eigenstates of (super)translation generators. 
In other words, the bulk dual of a simpleton at null infinity cannot be an ordinary field on Minkowski spacetime with a Fourier decomposition in plane waves (see also \cite{Ponomarev:2021xdq} for an earlier discussion of this issue in four bulk spacetime dimensions).

This is to be contrasted with the standard Wigner representations of the Poincar\'e (or BMS) algebra, in particular, with the faithful unitary irreducible representation of \mbox{$\mathfrak{iso}(d+1,1)$} corresponding to a massless scalar field on Minkowski spacetime $\mathbb{R}^{d+1,1}$. 
The latter representation space will be denoted $\mathcal D^{\mathfrak{iso}(d+1,1)}(\tfrac{d}2,0)$ for later purpose. For $d=2$, it was discussed as $\mathfrak{bms}_4-$module by Sachs in his seminal paper \cite{Sachs:1962wk}. 
Its higher-dimensional generalisation was described in \cite{Bekaert:2022ipg} in terms of scalar Carrollian primary fields $\varphi(u,\mathbf{x})$ on $\mathscr{I}_{d+1}$ with scaling dimension $d/2$ on which (super)translations act faithfully via the generators $f(\mathbf{x})\,\partial_u$.
We summarise the previous remarks in Tables \ref{tableprops} and \ref{tablesachs}. 

Upon restricting the Poincar\'e (or BMS) algebra to its Lorentz subalgebra $\mathfrak{so}(d+1,1)$,
the zero-(super)momentum representation remains irreducible and corresponds to the faithful representation of $\mathfrak{so}(d+1,1)$ denoted as $\mathcal{D}(\frac{d-1}2,0)$, which belongs to the complementary series of unitary irreducible representations of the Lorentz algebra.
On the contrary, upon restriction of the Poincar\'e (or BMS) algebra to the Lorentz subalgebra, the simpleton decomposes into the direct sum of two irreducible representations $\mathcal{D}(\Delta_\pm,0)$ spanned by the conformal primary scalar fields $\varphi_\pm(\mathbf x)$ on $S^d$ (with respective scaling dimension $\Delta_\pm=\frac{d\pm1}2$) and their descendants. More explicitly, this corresponds to the decomposition
\eqref{gensol}. 
This branching rule can be written as
\be\label{branchingflat}
\mathfrak{iso}(d+1,1)\downarrow\mathfrak{so}(d+1,1)\,:\quad \text{Simpleton}\,\downarrow\,\mathcal{D}\left(\tfrac{d-1}2,0\right)\,\oplus\,\mathcal{D}\left(\tfrac{d+1}2,0\right) .
\ee
However, the action of (super)translations is triangular, cf.~\eqref{uppertriangular} and \eqref{lowertriangular}, hence the simpleton is an indecomposable representation of Poincar\'e (and BMS) algebra. We summarise these remarks in Table \ref{tablerestrict}.

\begin{center}
\begin{table}[ht]
    \centering
    \renewcommand*{\arraystretch}{1.7}
    \begin{tabular}{c|c|c||c|c|c}
    Carroll & Poincar\'e & (Super)translation & Faithful & Unitary & Irreducible \\[-10pt]
    module & module & generators & & & \\
    \hline\hline
    Zero-energy & Zero-momentum & Trivial& No & Yes  & Yes \\\hline
    Simpleton & Simpleton  & Nilpotent & Yes & No & No   \end{tabular}
\caption{The two relevant modules of Carroll and Poincar\'e algebras. The first line is a submodule of the second line. \label{tableprops}}
\end{table}
\end{center}

\begin{center}
\begin{table}[ht]
    \centering
    \renewcommand*{\arraystretch}{1.7}
    \begin{tabular}{c|c|c||c|c|c}
    Poincar\'e & BMS & (Super)translation & Faithful & Unitary & Irreducible \\[-10pt]
    module & module & generators & & & \\
    \hline\hline
    Massless & Sachs & Diagonalisable & Yes & Yes  & Yes\\ \end{tabular}
\caption{The usual scalar massless representation of the Poincar\'e and BMS algebras, to be contrasted with Table \ref{tableprops}. \label{tablesachs}}
\end{table}
\end{center}

\begin{center}
\begin{table}[ht]
    \centering
    \renewcommand*{\arraystretch}{1.7}
    \begin{tabular}{c|c|c||c|c|c}
    Module of & Faithful & Irreducible & Module of & Faithful & Irreducible\\[-10pt]
    $\mathfrak{iso}(d+1,1)$ & & & $\mathfrak{so}(d+1,1)$ & &\\
    \hline\hline
    Zero-momentum & No & Yes & $\mathcal{D}(\frac{d-1}2,0)$ & Yes & Yes \\\hline
    Simpleton & Yes & No & $\mathcal{D}(\frac{d-1}2,0)\oplus\mathcal{D}(\frac{d+1}2,0)$ & Yes & No 
  \end{tabular}
  \caption{The two modules upon restriction of Poincar\'e algebra to its Lorentz subalgebra\label{tablerestrict}}
\end{table}
\end{center}

Being a free theory, the actions \eqref{eq:boundary-electric} and \eqref{eq:boundary-magnetic} of the simpleton are also invariant under additional `higher-spin' symmetries, described by products of Carrollian isometries with the composition as associative product, as shown in \cite{Bekaert:2022oeh} for the electric case and in Appendix~\ref{sec:symmetries-magnetic} for the magnetic case. 
Since these two theories are equivalent, it is not surprising that their (higher) symmetries close on the same algebra, that we denote as $\mathfrak{hsbms}^+_{d+2}$. 
The latter contains as a subalgebra the flat-space higher-spin algebra obtained in \cite{Campoleoni:2021blr} as a contraction of the infinite-dimensional algebra underlying Vasiliev's equations in AdS.\footnote{The link with the flat-space higher-spin algebra of \cite{Campoleoni:2021blr} can be understood by noticing that the latter results from the factorisation of an ideal generated by the condition $P_aP_b \sim 0$ in the universal enveloping algebra of $\mathfrak{iso}(d+1,1)$ (together with additional conditions involving the Lorentz generators). 
As we discussed, (super)translations are realised as nilpotent operators on the space of solutions \eqref{gensol}, and the other conditions defining the higher-spin algebra are also automatically realised on it \cite{Bekaert:2022ipg}.} 
A suitable gauging of this higher-spin algebra then reproduces the free equations of motion of an infinite tower of higher-spin fields in Minkowski spacetime \cite{Boulanger:2023prx}, thus suggesting a holographic interpretation also for the additional symmetries of the simpleton.

Let us also recall that the Poincar\'e algebra $\mathfrak{iso}(d+1,1)$ can be seen as the isometry algebra of both Minkowski spacetime $\mathbb R^{d+1,1}\cong \frac{ISO(d+1,1)}{SO(d+1,1)}$ and of AdS-Carroll spacetime $\mathbb R \times H_{d+1}\cong \frac{ISO(d+1,1)}{ISO(d+1)}$ (where AdS-Carroll\footnote{The AdS-Carroll spacetime is also referred to as `para-Minkowski' in the old terminology of \cite{Bacry:1968zf} or as the witticism `Poincarroll' in some private circles.} stands for the $c \to 0$ contraction of AdS space; see, e.g., \cite{Figueroa-OFarrill:2018ilb, Herfray:2021qmp}). 
The first spacetime is a Lorentzian manifold while the second one is a Carrollian manifold. 
In fact, these spacetimes are both homogeneous spaces of the Poincar\'e group, but quotiented respectively by Lorentz transformations or by the union of rotations and Carrollian boosts. 
Furthermore, $\mathscr I_{d+1}$ is a homogeneous space of the Poincar\'e group and can be seen as the codimension-one conformal boundary of both Minkowski and AdS-Carroll spacetimes. 
This will motivate our coming discussion on the bulk realisation of conformal Carrollian scalars on $\mathscr I_{d+1}$, as scalar fields living in Minkowski spacetime (in Section~\ref{sec:Minkowski}) or in AdS-Carroll spacetime (in Section~\ref{sec:ambient}).

In the following, we will also show that the simpleton representation can be obtained as a limit of Dirac's singleton in any spacetime dimensions, and this can be done in two ways. 
In Section~\ref{sec:limit-field}, we will first recover the general asymptotic analysis of free Minkowski scalar fields of Section~\ref{sec:Minkowski} as a limit of the corresponding analysis in AdS.
In Section~\ref{sec:singleton} we will then revisit this limit in purely algebraic terms, after reviewing the key properties of Dirac's singleton representation.

%%%%%%%%%%%%%%%%%%%%%%%%%%%%%%%%%%%%
\section{Bulk description on Minkowski spacetime} \label{sec:Minkowski}

Inspired by the standard description \cite{Fronsdal:1978vb, Angelopoulos:1980wg} of Dirac's singleton as a shortened AdS scalar field (which we will review in Section~\ref{sec:limit}), we show in this section that the simpleton can also be realised as a shortened Minkowski scalar. 
To this end,  we will first discuss in Section~\ref{sec:flat-scalar} the solution space of the Klein-Gordon equation for a massless scalar in Minkowski spacetime in terms of asymptotic data at null infinity, emphasising the crucial role of the scaling dimension. 
In Section~\ref{sec:holographic-simpleton}, we will then focus on the realisation of the simpleton as a shortened scalar with a specific scaling dimension, corresponding to that of a singleton in AdS. 
In Section~\ref{sec:exotic}, we will eventually show how the boundary data describing the simpleton can be rearranged in an alternative bulk description involving homogeneous fields in Minkowski spacetime, which makes manifest the nilpotency of (super)translation generators. 

%%%%%%%%%%%%%%%%%%
\subsection{On-shell scalars at null infinity} \label{sec:flat-scalar}

Recall that in Bondi/Eddington-Finkelstein coordinates $(r,u,\mathbf x)$, the Minkowski metric reads
\begin{equation}
    {\rm d}s_{\mathbb{R}^{d+1,1}}^2 = -\,{\rm d}u^2 - 2\,{\rm d}u\,{\rm d}r + r^2 {\rm d}\Omega_d^2 \, .
\end{equation}
We consider a massless scalar field $\upphi$ on Minkowski spacetime, admitting the following asymptotic expansion in these coordinates
\begin{equation}\label{1/rexp}
    \upphi(r,u,\mathbf x) = \frac{1}{r^\Delta} \sum_{n \geqslant 0} \frac{\phi_n(u,\mathbf x)}{r^n}\,,
\end{equation}
where $\phi_0$ is not identically vanishing and $\Delta$ is an arbitrary real number for the moment.
With this ansatz, the d'Alembert equation $\Box_{{}_{\mathbb R^{d+1,1}}}\upphi=0$ becomes equivalent to the following recursive set of equations for the coefficients $\phi_n$:
\begin{equation} \label{eq:recursion-Minkowski}
    (2\Delta + 2n-d) \dot \phi_n + \left[ \nabla_{S^d}^2 + (\Delta+n-1)(\Delta-d+n) \right] \phi_{n-1} = 0 \,.
\end{equation}
A detailed analysis of this system of equations can be found in \cite{Satishchandran:2019pyc, Bekaert:2022ipg}, and it has been extended in \cite{Campoleoni:2020ejn} to massless fields of arbitrary spin. Here we revisit these analyses stressing the group-theoretical structure of the solution space.

\paragraph{Solution space in the generic case ($\frac{d}2 - \Delta \notin \mathbb N$):} For asymptotic expansions such that\footnote{The equations \eqref{eq:recursion-Minkowski} have been analysed in \cite{Bekaert:2022ipg} in the exceptional cases $\Delta -\frac{d}{2}\in\mathbb N$ that correspond to the ambient formulation of Wick-rotated scalar singletons (`WRac') and their higher-order generalisations (described by GJMS operators), while the $\Delta = \tfrac{d}{2}$ case was already discussed in \cite{Satishchandran:2019pyc}.}  $\frac{d}2 - \Delta \notin \mathbb N$, the equations \eqref{eq:recursion-Minkowski} fix the expression of $\phi_n$ as a function of $\phi_{n-1}$ up to an `integration constant'
\begin{equation}\label{integrationconstant}
\psi_n(\mathbf x):=\phi_n(u=0,\mathbf x)\,,
\end{equation}
and the solution space identifies with this collection of integration constants. 
It is thus given by an infinite collection of arbitrary functions $\psi_n(\mathbf x)$ of the celestial sphere or, equivalently, by the value 
\begin{equation}\label{fctlightcone}
\uppsi(r,\mathbf x):=\upphi(r,u=0,\mathbf x) = \sum_{n \geqslant 0}r^{-(\Delta+n)} \psi_n(\mathbf x)    
\end{equation}
of the scalar field on the null cone through the origin (defined by the equation $u=0$ in Bondi/Eddington-Finkelstein coordinates).

The vector space of solutions to the d'Alembert equation with asymptotic expansion \eqref{1/rexp} is an infinite-dimensional $\mathfrak{iso}(d+1,1)-$module, which will be denoted 
\be\label{Viso}
\mathcal V^{\mathfrak{iso}(d+1,1)}(\Delta,0)\,:=\,\bigg\{\upphi(r,u,\mathbf x) = r^{-\Delta} \sum_{n \geqslant 0} r^{-n}\phi_n(u,\mathbf x)\,:\,\Box_{{}_{\mathbb{R}^{d+1,1}}}\upphi=0\bigg\}\,.
\ee
For any $N \in\mathbb N$,
one may consider the invariant subspace \be
\mathcal V^{\mathfrak{iso}(d+1,1)}(\Delta+N,0)\subset \mathcal V^{\mathfrak{iso}(d+1,1)}(\Delta,0)
\ee
of solutions of the form \eqref{1/rexp} where the coefficients $\phi_n(u,\mathbf x)$ vanish for $n=0,1,\ldots, N-1$.
In fact, the $\mathfrak{iso}(d+1,1)-$module \eqref{Viso} is indecomposable and there is a natural descending filtration by $\mathfrak{iso}(d+1,1)-$submodules, 
\be\label{filtration}
\mathcal V^{\mathfrak{iso}(d+1,1)}(\Delta,0)\supset\mathcal V^{\mathfrak{iso}(d+1,1)}(\Delta+1,0)\supset\mathcal V^{\mathfrak{iso}(d+1,1)}(\Delta+2,0)\supset\cdots
\ee

\paragraph{Leading-order quotient:} The previous filtration shows that one can quotient the solution space by any of the submodules appearing in \eqref{filtration}. Equivalently, this amounts to set to zero all $\psi_n(\mathbf{x})$ boundary data from a certain value of $n$ onwards. 
For instance, the quotient
\be \label{leading-quotient}
\mathcal W^{\mathfrak{iso}(d+1,1)}(\Delta,0):= \mathcal V^{\mathfrak{iso}(d+1,1)}\left(\Delta,0\right) \,\big /\, \mathcal V^{\mathfrak{iso}(d+1,1)}\left(\Delta+1,0\right)\quad \text{with}\quad \Delta\in\mathbb R\,,
\ee
is spanned by the leading boundary data 
\be
\phi_0(u,\mathbf x)=\lim\limits_{r\to\infty}\left[r^\Delta\upphi(r,u,\mathbf x)\right]
\ee
of solutions in \eqref{Viso}.

\paragraph{Radiative case ($\Delta = \frac{d}2$):} Often, the asymptotic analysis is restricted to the boundary data of radiative modes, which correspond to the exceptional value $\Delta=\tfrac{d}2$ for which the quotient
$\mathcal W^{\mathfrak{iso}(d+1,1)}(\tfrac{d}2,0)$ is spanned by arbitrary functions $\phi_0(u,\mathbf x)$ at null infinity, thanks to the vanishing of the coefficient in front of $\dot{\phi}_0$ in \eqref{eq:recursion-Minkowski}. 
This irreducible \mbox{$\mathfrak{iso}(d+1,1)-$}module was discussed as a $\mathfrak{bms}_{d+2}-$module by Sachs in his seminal paper \cite{Sachs:1962wk} (for $d=2$). 
As one can see, it is natural to identify the Sachs module with a quotient of Poincar\'e modules
\begin{equation}\label{DSachs}
    \text{Sachs}\,:\quad\mathcal D^{\mathfrak{iso}(d+1,1)}(\tfrac{d}2,0)\, :=\, \mathcal V^{\mathfrak{iso}(d+1,1)}(\tfrac{d}2,0)\, /\,\, \mathcal V^{\mathfrak{iso}(d+1,1)}(\tfrac{d+2}2,0)\,.
\end{equation}

\paragraph{Structure of the generic case ($\frac{d}2 - \Delta \notin \mathbb N$):} For the generic case $\Delta\neq \tfrac{d}2$, the leading-order quotient
$\mathcal W^{\mathfrak{iso}(d+1,1)}(\Delta,0)$ defined in \eqref{leading-quotient} is spanned by a conformal primary field $\phi_0(\mathbf x)$ on the celestial sphere $S^d$ of scaling dimension $\Delta$, together with all its descendants.
Let $\mathcal{V}^{\mathfrak{so}(d+1,1)}(\Delta,0)$ denote the (generalised) Verma module of \mbox{$\mathfrak{so}(d+1,1)$} spanned by scalar conformal primary fields on the celestial sphere $S^d$,  on which the (super)translations act trivially (since they do not depend on retarded time $u$). Therefore, one has the isomorphism of \mbox{$(\mathfrak{i})\mathfrak{so}(d+1,1)-$}modules
\be
\mathcal W^{\mathfrak{iso}(d+1,1)}\left(\Delta,0\right)\,\cong\,\mathcal V^{\mathfrak{so}(d+1,1)}(\Delta,0)\quad\text{for}\quad\Delta\neq\tfrac{d}2\,.
\ee
Consequently, for $\frac{d}2 - \Delta \notin \mathbb N$, all the quotients $\mathcal W^{\mathfrak{iso}(d+1,1)}(\Delta+n,0)$ for non-negative integer $n\in\mathbb N$ are spanned by the `integration constants' \eqref{integrationconstant} which are conformal primary fields on the celestial sphere $S^d$ of scaling dimension $\Delta+n$.
The graded vector space associated to the descending filtration \eqref{filtration} is, by definition, the direct sum
\be\label{gradedspace}
\text{gr}\,\mathcal V^{\mathfrak{iso}(d+1,1)}(\Delta,0)\,=\,\bigoplus_{n=0}^\infty
\mathcal W^{\mathfrak{iso}(d+1,1)}(\Delta+n,0)\,.
\ee
Therefore, we have the isomorphism of $(\mathfrak{i})\mathfrak{so}(d+1,1)-$modules
\be\label{isomgraded}
\text{gr}\,\mathcal V^{\mathfrak{iso}(d+1,1)}(\Delta,0)\,\cong\,\bigoplus_{n=0}^\infty
\mathcal V^{\mathfrak{so}(d+1,1)}(\Delta+n,0)\quad\text{for}\quad \frac{d}2 - \Delta \notin \mathbb N \,.
\ee
In more concrete terms, this corresponds to Equation \eqref{fctlightcone} and the corresponding discussion above.

\paragraph{Further quotients:} More generally, one may consider quotients
\be\label{quotientN}
\mathcal V^{\mathfrak{iso}(d+1,1)}\left(\Delta,0\right) \,\big /\, \mathcal V^{\mathfrak{iso}(d+1,1)}\left(\Delta+N,0\right)
\ee
for integers $N\geqslant1$. They are $\mathfrak{iso}(d+1,1)-$modules spanned by equivalence classes of solutions of d'Alembert equation, $\upphi(r,u,\mathbf x)$, modulo terms ${\cal O}(1/r^{\Delta+N})$.
Note that for $\frac{d}2 - \Delta \notin \mathbb N$, such quotient modules are isomorphic to direct sums 
\begin{eqnarray}
&&\mathfrak{iso}(d+1,1)\downarrow\mathfrak{so}(d+1,1)\,:\nonumber\\
&&\qquad\mathcal V^{\mathfrak{iso}(d+1,1)}\left(\Delta,0\right) \,\big /\, \mathcal V^{\mathfrak{iso}(d+1,1)}\left(\Delta+N,0\right)=
\bigoplus_{n=0}^{N-1}
\mathcal V^{\mathfrak{so}(d+1,1)}(\Delta+n,0) \, .
\label{orderNquotient}
\end{eqnarray}
As one will see in the next subsection, the simpleton corresponds to the case $\Delta=\tfrac{d-1}2$ and $N=2$.

%%%%%%%%%%%%%%%%%%
\subsection{Holographic realisation as shortened scalar} \label{sec:holographic-simpleton}

For the particular value $\Delta = \Delta_- = \frac{d-1}{2}$, the first two equations in the recursion relations~\eqref{eq:recursion-Minkowski} are
\begin{equation}
    \dot\phi_0 = 0 \,,\quad \dot \phi_1 + \hat\nabla^2 \phi_0 = 0 \,.
\end{equation}
As already noticed in \cite{Figueroa-OFarrill:2023qty}, they are precisely the equations of motion \eqref{eq:boundary-magnetic-eom} in the magnetic formulation of the simpleton, where $\hat\nabla^2$ was defined in \eqref{shiftedLaplacian}.
It is therefore natural to identify the simpleton with the quotient of Poincar\'e modules
\begin{equation}\label{Dsimpleton}
    \text{Simpleton}\,:\quad\mathcal D^{\mathfrak{iso}(d+1,1)}(\tfrac{d-1}2,0)\, :=\, \mathcal V^{\mathfrak{iso}(d+1,1)}(\tfrac{d-1}2,0)\, /\,\, \mathcal V^{\mathfrak{iso}(d+1,1)}(\tfrac{d+3}2,0)\,.
\end{equation}
The quotient on the right-hand side of \eqref{Dsimpleton} means that the boundary (magnetic) description of the on-shell simpleton on null infinity $\mathscr{I}_{d+1}$ is equivalent to the space of solutions of a massless scalar field on Minkowski spacetime $\mathbb{R}^{d+1,1}$ which admits an asymptotic expansion (in power series of $1/r$) with leading term $1/r^{\frac{d-1}2}$ and modulo solutions of order $1/r^{\frac{d+3}2}$, that is to say
\begin{equation}
\Box_{{}_{\mathbb{R}^{d+1,1}}}\upphi=0\,,\quad \upphi(r,u,\mathbf x)=\frac{\phi_0(u,\mathbf x)}{r^{\frac{d-1}2}}+\frac{\phi_1(u,\mathbf x)}{r^{\frac{d+1}2}}\quad\text{modulo}\;\;{\cal O}\left(\frac1{\,r^{\frac{d+3}2}\,}\right) .
\end{equation}

With the identification \eqref{Dsimpleton} for the simpleton, the branching rule \eqref{branchingflat} can be written as the following equality of $\mathfrak{so}(d+1,1)-$modules
\begin{eqnarray} \label{eq:bulk-simpleton}
&&\mathfrak{iso}(d+1,1)\downarrow\mathfrak{so}(d+1,1)\,:\nonumber\\[5pt]
&&\qquad\mathcal{D}^{\mathfrak{iso}(d+1,1)}(\tfrac{d-1}2,0)=\mathcal{D}^{\mathfrak{so}(d+1,1)}(\tfrac{d-1}2,0)\oplus\mathcal{D}^{\mathfrak{so}(d+1,1)}(\tfrac{d+1}2,0)\,.
\end{eqnarray}
%

%%%%%%%%%%%%%%%%%%
\subsection{Doublet realisation as homogeneous fields\label{sec:exotic}}

One way of selecting a representative in the previous quotient from the holographic description of the simpleton is to work with fields which are homogeneous in the Cartesian coordinates on Minkowski spacetime.

Indeed, the Euler operator $x^a \partial_a$ ($a=0,1,\dots,d+1$) on $\mathbb{R}^{d+1,1}$ reduces to the operator $r\,\partial_r + u\, \partial_u$ in Bondi/Eddington-Finkelstein coordinates.  Let us consider again a free massless scalar field $\upphi(r,u,\mathbf x)$ with asymptotic expansion
\eqref{1/rexp} such that $\frac{d}2 - \Delta \notin \mathbb N$. 
Then, the equations \eqref{eq:recursion-Minkowski} can be solved order-by-order by specifying, for all $n\in\mathbb N$, the `integration constants' $\psi_n(\mathbf x)$, which are functions on the celestial sphere defined by \eqref{integrationconstant}.\footnote{In order to solve the Cauchy problem on $\mathscr I_{d+1}$, the integration constants could be specified at any arbitrary $u=u_0$, but $u_0=0$ plays a special role since, as detailed in the following, in this section we work with fields homogeneous in the Cartesian coordinates.} 
At order $r^{-\Delta - n}$, the field $\phi_n(u,\mathbf x)$ is polynomial in $u$ of order $n$,
\begin{equation}\label{polynomialexp}
\phi_n(u,\mathbf x)=\sum\limits_{p=0}^n u^{n-p}\,\mathcal{D}_{2(n-p)}\psi_{p}(\mathbf x) \,,
\end{equation}
with coefficients that are obtained by applying a linear differential operator $\mathcal{D}_{2k}$ on the celestial sphere of order $2k$ on the `integration constant' $\psi_{n-k}\,$ (see, e.g., \cite{Bekaert:2022ipg} for details). 
In particular, $\phi_n(u=0,\mathbf x)=\psi_{n}(\mathbf x)\,$.
Therefore, combining \eqref{1/rexp} and \eqref{polynomialexp} implies that the scalar field takes the form
\begin{equation}\label{formalexp}
\upphi(r,u,\mathbf x) =  \sum\limits_{p,q=0}^\infty \frac{1}{r^{\Delta+p}}\left(\frac{u}{r}\right)^q \mathcal{D}_{2q}\psi_{p}(\mathbf x)\,.
\end{equation}
As one can see, the on-shell fields $\upphi$ of the form \eqref{1/rexp} admit a decomposition
\begin{equation}
\upphi(r,u,\mathbf x) =  \sum\limits_{p=0}^\infty \upphi_{[p]}(r,u,\mathbf x)\,,\end{equation}
in fields $\upphi_{[p]}$ which are homogeneous functions in $r$ and $u$ of degree $p\in\mathbb N$,
\begin{equation} \label{eq:diff-op}
\upphi_{[p]}(r,u,\mathbf x):= \frac{1}{r^{\Delta+p}}\,\chi_p(\tfrac{u}{r},\mathbf x)\quad\text{and}\quad\chi_p(z,\mathbf x):=\sum\limits_{q=0}^\infty z^q \,\mathcal{D}_{2q}\psi_{p}(\mathbf x)\,.
\end{equation}
Switching to Cartesian coordinates, this means that the
field $\upphi(x)$ can be decomposed as a sum of terms $\upphi_{[n]}(x)$ with homogeneity $\Delta + n$ in the Cartesian coordinates $x^a$ (\mbox{$a=0,1,\ldots,d+1$}), i.e.
\begin{equation}\label{Ndecomp}
    \upphi(x)=  \sum_{n=0}^\infty \upphi_{[n]}(x) \,,\quad \text{with}\quad\left(x^a \partial_a + \Delta + n\right) \upphi_{[n]}(x) = 0 \,.
\end{equation}
This decomposition in homogeneity degree implicitly assumes a choice of origin and corresponds to a choice of representative in the quotient spaces $\mathcal W^{\mathfrak{iso}(d+1,1)}(\Delta+n,0)$. In other words, the decompositions \eqref{formalexp} and \eqref{Ndecomp} provide an explicit description of the graded vector space \eqref{gradedspace}.

Note that $\Box_{{}_{\mathbb R^{d+1,1}}}\upphi=0$ implies $\Box_{{}_{\mathbb R^{d+1,1}}} \upphi_{[n]}=0$ for each $n\in\mathbb N$, since the d'Alembertian operator has a fixed homogeneity degree (equal to $-2$). In other words, one can reformulate the problem of looking for solutions of the d'Alembert equation $\Box_{{}_{\mathbb R^{d+1,1}}}\upphi=0$ satisfying the asymptotic expansion \eqref{1/rexp} into the problem of looking for the solutions to an infinite system of decoupled equations 
\be\label{exoticsystem}
\Box_{{}_{\mathbb{R}^{d+1,1}}}\upphi_{[n]}=0\,,\qquad\left(x^a \partial_a + \Delta + n\right) \upphi_{[n]} = 0 \,,
\ee
which has the virtue of making the Lorentz invariance manifest, while it is hidden in the infinite system \eqref{eq:recursion-Minkowski} of coupled equations. However, the equivalent formulation \eqref{exoticsystem} is somewhat exotic because the translation invariance is not manifest since it involves a choice of origin in Minkowski spacetime. 
Nevertheless, the whole system \eqref{exoticsystem} of equations is of course Poincar\'e invariant but the action of translations is not diagonal: \mbox{$\delta\upphi_{[n]}=\epsilon^a\partial_a\upphi_{[n+1]}$} since the translation generator has homogeneity degree $-1$. 
Readers familiar with the embedding approach to conformal geometry will recognise that \eqref{exoticsystem} provides the ambient formulation of a conformal primary field on the celestial sphere $S^d$ of scaling dimension $\Delta + n$. 
Therefore, the decomposition \eqref{Ndecomp} provide an explicit realisation of the isomorphism  \eqref{isomgraded}.

Moreover, consistent truncations, to be understood as quotients \eqref{quotientN} of Poincar\'e modules, can be performed by setting some $\upphi_{[n]}$ to zero (even an infinite number of them, which amounts to work only with $\left\{ \upphi_{[n]} \right\}_{0 \leqslant n \leqslant N}$ for some $N$). 
This choice is consistent, since each $\upphi_{[n]}$ can be expressed in terms of the action of a differential operator acting on a single integration constant $\psi_n(\mathbf x)$, see \eqref{eq:diff-op}.

\paragraph{Leading-order quotient in the non-radiative case} For $N = 0$ and $\Delta \neq \frac{d}{2}$, we are modelling a bulk scalar field with only one arbitrary boundary integration constant $\psi_0$.
This corresponds to a massless Carroll particle (i.e.\ a zero-energy representation of the Carroll group) or, equivalently, to the Poincar\'e module $\mathcal W^{\mathfrak{iso}(d+1,1)}(\Delta,0)$ where translations are realised trivially (i.e.\ a zero-momentum representation of Poincar\'e group described in Tables~\ref{tableprops} and \ref{tablerestrict}).

\paragraph{Subleading-order quotient corresponding to the simpleton} The case of the simpleton corresponds to $\Delta = \frac{d-1}{2}$ and $N = 1$. 
It can be formulated in terms of a pair of bulk fields $\upphi_-=\upphi_{[0]}$ and $\upphi_+=\upphi_{[1]}$
with homogeneity degree $\Delta_\mp=\tfrac{d\mp 1}2$ in Cartesian coordinates
\be \label{eq:exotic-doublet}
\Box_{{}_{\mathbb R^{d+1,1}}}\upphi_\pm(x)=0\,,\qquad \left(x^a \partial_a + \tfrac{d\pm 1}2 \right) \upphi_\pm(x) = 0 \,,
\ee
The generators of translations and Lorentz transformations acting on the doublet
\be\label{doublet}
\boldsymbol{\upphi}(x) = \begin{pmatrix} \upphi_-(x) \\ \upphi_+(x) \end{pmatrix} ,
\ee
read
\be
P_a =
\begin{pmatrix}
0 & 0 \\ \partial_a & 0
\end{pmatrix} ,\quad
J_{ab} =
\begin{pmatrix}
2x_{[a}\partial_{b]} & 0 \\ 0 & 2x_{[a}\partial_{b]}
\end{pmatrix}.
\ee
Finally, we display here for convenience the closed form expressions for $\upphi_\pm(r,u,\mathbf x)$
\begin{align} \label{eq:fields-scri}
\upphi_-(r,u,\mathbf x) &= \frac{1}{r^{\Delta_-}} \sum_{n \geqslant 0} \left(\frac{2u}{r}\right)^n \frac{1}{(2n)!} \prod_{k=0}^{n-1} \left(\Delta_-^2 - \nabla^2 - k^2\right) \psi_-(\mathbf x) \,,\\
\upphi_+(r,u,\mathbf x) &= \frac{1}{r^{\Delta_+}} \sum_{n \geqslant 0} \left(\frac{2u}{r}\right)^n \frac{1}{(2n+1)!} \prod_{k=1}^n \left(\Delta_-^2-\nabla^2 - k^2\right) \psi_+(\mathbf x) \,,
\end{align}
where $\psi_-(\mathbf x) = \psi_0(\mathbf x)$ and $\psi_+(\mathbf x) = \psi_1(\mathbf x)$ are the two integration constants spanning the solution space of the magnetic theory and forming the Poincar\'e-module $\mathcal D^{\mathfrak{iso}(d+1,1)}(\frac{d-1}{2},0)$.

%%%%%%%%%%%%%%%%%%%%%%%%%%%%%%%%%%%%
\section{From a flat/Carrollian limit of the singleton}
\label{sec:limit}

In this section, we review the properties of Dirac's singleton, first looking at its realisation in terms of the space of solutions of the AdS Klein-Gordon equation with a specific mass and then in terms of its definition as an irreducible representation of the AdS isometry algebra. 
This will allow us to show that the previous bulk and group-theoretical descriptions of the simpleton can be recovered by taking a suitable limit of the corresponding descriptions of Dirac's singleton.

%%%%%%%%%%%%%%%%%%
\subsection{Field-theoretical description}
\label{sec:limit-field}

Consider the AdS$_{d+2}$ metric in Bondi/Eddington-Finkelstein coordinates, which can be obtained from global coordinates by switching to radial tortoise coordinates:
\begin{equation}
    {\rm d}s_{\text{AdS}_{d+2}}^2 = -\left(1+ \frac{r^2}{R^2}\right) {\rm d}u^2 - 2\,{\rm d}u\,{\rm d}r + r^2 {\rm d}\Omega_d^2 \, ,
\end{equation}
where $R$ denotes the AdS curvature radius.
In this coordinate system, the Klein-Gordon equation 
$\left( \nabla^2_\text{AdS$_{d+2}$} -m^2 \right) \upphi = 0$
reads
\begin{equation}\label{klein-gordon}
\left[ \left(1+\frac{r^2}{R^2}\right) \partial_r{}^2 - 2\,\partial_r \partial_u + \left(\frac{d}{r}+(d+2)\frac{r}{R^2}\right)  \partial_r - \frac{d}{r}\, \partial_u  + \frac{1}{r^2} \nabla_{S^d}^2 -m^2 \right] \upphi = 0 \, .
\end{equation}

We assume again that asymptotically the scalar field $\upphi$ can be expanded in a power series of the null radial coordinate $r$ starting at a power $\Delta$, as in \eqref{1/rexp}.
The Klein-Gordon equation \eqref{klein-gordon} then implies, at order $r^{-(\Delta+n+1)}$,
\begin{equation} \label{recurrence}
\begin{split}
   & (2\Delta + 2n-d) \dot \phi_n + \left[ \nabla_{S^d}^2 + (\Delta+n-1)(\Delta-d+n) \right] \phi_{n-1} \\
   &\qquad+ \frac{1}{R^2} \left[(\Delta+n+1)(\Delta+n-d)-(mR)^2 \right] \phi_{n+1} = 0 \,.
\end{split}
\end{equation}

For $n = -1$, since $\phi_0 \neq 0$, we get the standard mass-shell relation between the mass $m$ and the asymptotic scaling dimension $\Delta$:
\begin{equation}\label{masssquared}
    m^2 = -\frac{\Delta(d+1-\Delta)}{R^2} \,.
\end{equation}
From now on, we will assume that this relation is satisfied. With this relation between $m$ and $\Delta$, the coefficient in \eqref{recurrence} in front of $\phi_{n+1}$ takes the expression $R^{-2}(n+1)(2\Delta + n-d)$.

For $n=0$, we get
\begin{equation} \label{eq:first-equation}
    (2\Delta-d)\left(\dot \phi_0 + \frac{1}{R^2} \phi_1\right) = 0 \,.
\end{equation}
Two cases should be distinguished:
\begin{itemize}
\item \underline{$\Delta = d/2$}: the previous equation \eqref{eq:first-equation} is trivially satisfied, meaning that $\phi_1$ is a priori unrelated to $\phi_0$.
This is the analogue of the radiative solutions in Minkowski spacetime.\footnote{Incidentally, note that the corresponding mass-squared $m^2=-d(d+2)/(2R)^2$ coincides with the values of the curvature term in the Yamabe operator of AdS$_{d+2}$. We are grateful to S.~I.~A.~Raj for pointing this out to us.} 
In other words, this case corresponds to the leading branch of solutions for a conformal scalar on AdS spacetime (the other branch has scaling dimension $\Delta\,=\,d/2\,+\,1 $).
\item \underline{$\Delta \neq d/2$}: the above equation \eqref{eq:first-equation} determines $\phi_1$ in terms of the retarded time derivative of $\phi_0$.\footnote{An explicit comparison between the parameterisation of the solution space in Bondi/Eddington-Finkelstein and Fefferman-Graham coordinates is given, e.g., in \cite{Campoleoni:2023eqp} for $\Delta = 0$ and $m=0$.}
\end{itemize}

For $n = 1$, we get
\begin{equation} \label{eq:second-equation}
    (2\Delta+2-d)\dot \phi_1 + \left[ \nabla_{S^d}^2 + \Delta(\Delta-d+1) \right] \phi_0 + \frac{2}{R^2} (2\Delta+1-d) \phi_2 = 0 \,.
\end{equation}
Here, we again see a qualitative difference depending on the value of $\Delta$. If $2\Delta+1-d= 0$ (i.e. $\Delta=\tfrac{d-1}2$, which corresponds to the scalar singleton on AdS$_{d+2}$), 
the factor in front of $\phi_2$ vanishes and after replacing in \eqref{eq:second-equation} the expression $\phi_1=-R^2\dot\phi_0$ (obtained at the previous order), we obtain an equation on $\phi_0$:\footnote{If one assumes an expansion in integer powers of $r$, as it is often the case in the literature, the coefficient in front of $\phi_2$ in \eqref{eq:second-equation} can only vanish in odd spacetime dimensions. 
In this case, at least in Fefferman-Graham coordinates, one usually introduces a logarithmic branch in the ansatz \eqref{1/rexp} that allows one to avoid imposing an equation on $\phi_0$.}
\begin{equation}
\label{LaplaceYamabe}
\phi_1=-R^2\dot\phi_0\,,\quad    \dot\phi_1 + \left[ \nabla_{S^d}^2 - \left(\tfrac{d-1}{2}\right)^2 \right] \phi_0 = 0 \quad\Longrightarrow\quad - R^2 \ddot\phi_0 + \hat\nabla^2 \phi_0 = 0 \,.
\end{equation}
This last equation is nothing but the relativistic Laplace-Yamabe equation for a scalar field $\phi_0$ on Einstein universe $\mathbb R \times S^d$, where the inverse AdS curvature radius $R^{-1}$ plays the role of an effective speed of light on the boundary (cf.\ Section~\ref{electric/magnetic}). 
For any other value of the scaling dimension, i.e. for $\Delta\neq\tfrac{d-1}2$, the field $\phi_2$ can be expressed algebraically in terms of $\phi_1$ and $\phi_0$.

Here, we see the special character of the singleton in that it is the only value of $\Delta$ for which the subleading branch of solutions starting from $\phi_2$ can be decoupled and factored out from the space of solutions of the Klein-Gordon equation. 
In fact, the standard bulk description \cite{Fronsdal:1978vb, Angelopoulos:1980wg} (see also \cite{Bekaert:2012ux, Bekaert:2012vt, Leigh:2012mz, Bekaert:2013zya} for a discussion in any dimension) of the singleton is as the space of solutions of a scalar field on AdS$_{d+2}$ spacetime with critical mass $m^2=\frac{(1-d)(d+3)}{(2R)^2}$ which admits an asymptotic expansion with leading term $r^{-\frac{d-1}2}$ and modulo solutions of order $r^{-\frac{d+3}2}$, that is to say
\begin{equation}
\left(\nabla^2_\text{AdS$_{d+2}$}-m^2\right)\upphi = 0\,,\quad
\upphi(r,u,\mathbf x)=\frac{\phi_0(u,\mathbf x)}{r^{\frac{d-1}2}}+\frac{\phi_1(u,\mathbf x)}{r^{\frac{d+1}2}} \;\text{ modulo }\;{\cal O}\left(\tfrac1{\,r^{\frac{d+3}2}\,}\right) .
\end{equation}
As one can see, the flat limit of the bulk singleton reproduces the bulk description \eqref{eq:bulk-simpleton} of the simpleton.
Note that the boundary data $\phi_0$ is \textit{on-shell} and can be interpreted as a dynamical conformal scalar field living on $\mathbb R \times S^d$.
Indeed, the standard boundary description of the singleton is as an on-shell conformal scalar field on  the boundary of AdS.
In this sense, the quotient in the bulk description of the singleton has the effect of removing all local bulk degrees of freedom. 
This representation of the AdS isometry group is extremely degenerate from a physical point of view. In fact, the product of two transvection generators does not vanish (contrarily to the bulk simpleton) but it almost vanishes, in the sense that it is proportional to a (sum of) product(s) of two Lorentz generators. 
We recover in this way standard facts about the singleton, usually obtained in a Fefferman-Graham expansion or ambient formulation (see, e.g., Section~2 of \cite{Bekaert:2013zya} for a review).

Taking directly the flat limit of \eqref{recurrence} gives back \eqref{eq:recursion-Minkowski}. Note that, in the flat limit, the mass vanishes in agreement with \eqref{masssquared} and the scaling dimension $\Delta$ is not related to the mass any more. 
Focusing on the singleton case $\Delta = \tfrac{d-1}2$, the flat limit $R \to \infty$ can be taken in two ways. On the one hand, one can take the limit in Eq.~\eqref{LaplaceYamabe}, obtained after replacing $\phi_1$ by its expression in terms of $\phi_0$. One obtains directly the electric theory $\ddot \phi = 0$ upon the identification $\phi=\phi_0$. On the other hand, keeping equations in first-order form and looking at the $R \to \infty$ limit of \eqref{eq:first-equation} and \eqref{eq:second-equation}, we obtain $\dot\phi_0 = 0$ and $\dot\phi_1 = - \hat\nabla^2 \phi_0$ which reproduces the magnetic theory \eqref{eq:boundary-magnetic-eom} upon the further identification $\pi=-\phi_1$, as explained in Section~\ref{sec:Minkowski}. 
This computation is the holographic counterpart of what we explained from an intrinsic point of view in Section~\ref{electric/magnetic} and is yet another proof that the electric and magnetic theories have the same common origin.

%%%%%%%%%%%%%%%%%%
\subsection{Group-theoretical description}\label{sec:singleton}

The standard group-theoretical description of the scalar singleton (also called `Rac') is as the `ultrashort' (in physicist jargon) or `minimal' (in mathematician jargon) representation of the algebra $\mathfrak{so}(d+1,2)$ (see, e.g., \cite{Kobayashi_2003} and references therein). 
Slightly more concretely, it is a lowest-weight irreducible module $\mathcal D^{\mathfrak{so}(d+1,2)}(\tfrac{d-1}2,0)$ of the conformal algebra, which is induced from the trivial representation of $\mathfrak{so}(d+1)$ (i.e.\ a `scalar field') and whose scaling dimension saturates the unitarity bound $\Delta\geqslant\tfrac{d-1}2$ when $s=0$.

Let $\mathcal{V}^{\mathfrak{so}(d+1,2)}(\Delta,0)$ denote the (generalised) Verma module of $\mathfrak{so}(d+1,2)$ spanned by a scalar conformal primary field on the conformal boundary $\mathbb{R}\times S^d$ of AdS$_{d+2}$, together with all its descendants. 
The scalar singleton is the following quotient of two \mbox{$\mathfrak{so}(d+1,2)-$}modules
\begin{equation}\label{Dsingleton}
    \text{Singleton}\,:\quad\mathcal D^{\mathfrak{so}(d+1,2)}\left(\tfrac{d-1}2,0\right)\, :=\, \mathcal V^{\mathfrak{so}(d+1,2)}\left(\tfrac{d-1}2,0\right) /\,\, \mathcal V^{\mathfrak{so}(d+1,2)}\left(\tfrac{d+3}2,0\right) .
\end{equation}
Upon restricting the AdS isometry algebra $\mathfrak{so}(d+1,2)$ to its Lorentz subalgebra $\mathfrak{so}(d+1,1)$, the branching rule of the singleton (see, e.g., \cite{Basile:2017kaz}) 
decomposes\footnote{We thank T.~Basile for pointing out to us the relevance of this branching rule, in analogy with the structure of the simpleton.} into the direct sum of two irreducible representations $\mathcal{D}(\Delta_\pm,0)$ spanned by conformal primary scalar fields on $S^d$ with respective scaling dimension $\Delta_\pm=\frac{d\pm1}2$,
\be
\mathfrak{so}(d+1,2)\downarrow\mathfrak{so}(d+1,1)\,:\quad \text{Singleton}\,\downarrow\,\mathcal{D}\left(\tfrac{d-1}2,0\right)\,\oplus\,\mathcal{D}\left(\tfrac{d+1}2,0\right) .
\ee
Phrased differently, in analogy with \eqref{eq:bulk-simpleton}, this branching rule can be written as the following equality of $\mathfrak{so}(d+1,1)-$modules
\be
\mathcal{D}^{\mathfrak{so}(d+1,2)}\left(\tfrac{d-1}2,0\right)=\mathcal{D}^{\mathfrak{so}(d+1,1)}\left(\tfrac{d-1}2,0\right)\oplus\mathcal{D}^{\mathfrak{so}(d+1,1)}\left(\tfrac{d+1}2,0\right) .
\ee

From the point of view of the conformal algebra $\mathfrak{so}(d+1,2)$ spanned by the generators $J_{\mu\nu}$ of Lorentz transformations, $D$ of dilations, $P_\mu$ of translations and $K_\mu$ of special conformal transformations, where \mbox{$\mu,\nu \in \{ 0, 1, \ldots, d \}$}, the generalised Verma module corresponding to the singleton is built upon the lowest-weight state $|\phi\rangle$ which is a scalar primary
\begin{equation}
    K_\mu |\phi\rangle = 0 \,,\quad J_{\mu\nu} |\phi\rangle = 0 \,,\quad D |\phi\rangle = \Delta |\phi\rangle \,.
\end{equation}
The associated generalised Verma module is built in the following way, where $\Delta = \frac{d-1}{2}$,
\begin{equation}\label{V-}
    \mathcal V^{\mathfrak{so}(d+1,2)}\left(\tfrac{d-1}{2},0\right) = \text{span}\left\{P_{\mu_1} \,\cdots\, P_{\mu_s}\,|\phi\rangle \right\}_{s \geqslant 0} \,.
\end{equation}
This module admits a submodule 
\begin{equation}\label{V+}
    \mathcal V^{\mathfrak{so}(d+1,2)}\left(\tfrac{d+1}{2},0\right) = \text{span}\left\{P_{\mu_1} \,\cdots\, P_{\mu_t}\,P^2\,|\phi\rangle \right\}_{t \geqslant 0} \,.
\end{equation}
which is spanned by contractions of $P_\mu$ or, equivalently, generated by the state $P^2|\phi\rangle$ (where $P^2 = P_\mu P^\mu$) which is both primary and descendant.
The quotient \eqref{Dsingleton} of the module \eqref{V-} by the submodule \eqref{V+}
makes the corresponding scalar field on-shell, giving rise to the singleton representation.

In order to make contact with the previous sections on the simpleton, one should first note that, in addition to the unitary irreducible representations of the Poincar\'e group classified by Wigner, one can also define lowest-weight (possibly non-unitary, see \cite{Bagchi:2019xfx, Chen:2021xkw, Bagchi:2022owq}) representations of the Poincar\'e algebra $\mathfrak{iso}(d+1,1)$, where the latter is seen as a conformal Carroll algebra, in close analogy with the lowest-weight representations of the conformal algebra $\mathfrak{so}(d+1,2)$. 
Furthermore, modules of the Poincar\'e algebra can be defined starting from modules of the conformal algebra by taking the limit \mbox{$c \rightarrow 0\,$}, as explained in \cite{Bagchi:2019xfx}. Representations that are lowest-weight and scalar remain lowest-weight and scalar in the sense that
\begin{equation}
K_i |\varphi\rangle = 0 \,,\quad K_0 |\varphi\rangle = 0 \,,\quad J_{ij} |\varphi\rangle = 0 \,,\quad B_i |\varphi\rangle = 0 \,,\quad D |\varphi\rangle = \Delta |\varphi \rangle \,,
\end{equation}
for the electric limit, where $B_i$ are Carrollian boosts (see Appendix~\ref{sec:limits-bulk-bdy} for conventions). For the singleton, the value $\Delta = \frac{d-1}{2}$ stays the same in the limit, and we can define the quotient \eqref{Dsimpleton} of induced $\mathfrak{iso}(d+1,1)-$modules in the same way as before.
This leads to
\begin{equation} \label{eq:UR-Verma}
\mathcal D^{\mathfrak{iso}(d+1,1)}\left(\tfrac{d-1}2,0\right) =\,\text{span}\left\{P_{i_1} \,\cdots\, P_{i_s}\,|\varphi\rangle \right\}_{s \geqslant 0} \ \oplus\ \text{span}\left\{P_{i_1} \,\cdots\, P_{i_t}\,P_0\,|\varphi\rangle \right\}_{t \geqslant 0} \,,
\end{equation}
where we discarded the product of more than one $P_0$ on $|\varphi\rangle$ in the universal enveloping algebra of $\mathfrak{iso}(d+1,1)$ by account of the Carrollian limit of the quotient condition \mbox{$P^2\,|\phi\rangle \sim 0$}, giving $P_0\,P_0\,|\varphi\rangle\sim 0$. 
We can readily see the emergence of a semi-direct structure, related to the presence of an Abelian factor generated by the action of the universal enveloping algebra of the conformal Carroll algebra on the second part of Eq.~\eqref{eq:UR-Verma}.

To summarise, in the In\"on\"u-Wigner contraction corresponding to the flat/Carrollian limit the singleton admits a smooth limit which is the simpleton:
\be
\text{Singleton} \stackrel{\mathfrak{so}(d+1,2)\,\to\,\mathfrak{iso}(d+1,1)}{\longrightarrow} \,\text{Simpleton} \,.
\ee
Notice that this refines the proposal of \cite{Flato:1978qz}, according to which the flat limit of the singleton should only correspond to the zero-momentum representation of the Poincar\'e algebra, which instead only corresponds to an invariant subspace within the simpleton representation (see also \cite{Ponomarev:2021xdq}).

%%%%%%%%%%%%%%%%%%%%%%%%%%%%%%%%%%%%
\section{Bulk description on AdS-Carroll spacetime} \label{sec:ambient}

In this section, we propose ambient-space descriptions of both the electric and the magnetic theories, allowing us to also describe them as the boundary data of bulk fields in AdS-Carroll spacetime. 
The latter data is associated to a \emph{shortened} ultra-relativistic massless scalar fields, i.e.\ to an exotic collection of modes that carry less degrees of freedom than the $c \to 0$ limit of a scalar field in AdS. From the boundary point of view, this corresponds to the fact that the boundary data correspond to an \emph{on-shell} Carrollian scalar field. 
This process is reminiscent of the shortening condition in AdS defining a \emph{singleton} from a massless scalar by tuning the scaling dimension of the leading order field in an asymptotic expansion (cf.~Section~\ref{sec:limit}).

The ambient formulation of the electric theory of the conformal Carrollian scalar field was already given in \cite{Bekaert:2022oeh}, while the bulk description of the electric theory and the corresponding ones for the magnetic theory are new.
The starting point to establish both the ambient and bulk descriptions of the simpleton is the ambient space description of the AdS singleton. 
We recall that the latter is usually given in terms of a triple of constraints (more precisely, two equations and one equivalence relation)
\begin{equation} \label{eq:ambient-relativistic}
    \Box_{{}_{\mathbb R^{d+1,2}}}\,\Phi = 0 \,,\quad \left(X^A \partial_A + \tfrac{d-1}{2}\right) \Phi = 0 \,,\quad \Phi \sim \Phi + X^2 \Psi \,,
\end{equation}
where the coordinates $X^A$ describe the ambient space $\mathbb R^{d+1,2}$ with flat metric $\eta_{AB}$ of signature $(-,+,\dots,+,-)$, of which AdS$_{d+2}$ is a slice of constant curvature $X^2 = -R^2$. 
The value $\frac{d-1}{2}$ is tuned such that the three equations, seen as constraints, are compatible. These constraints close on the $\mathfrak{sp}(2,\mathbb{R})$ algebra, with the commutator as Lie bracket. 
The constraint algebra $\mathfrak{sp}(2,\mathbb{R})$ and the isometry algebra $\mathfrak{so}(d+1,2)$ form a Howe dual pair, in the sense that they centralise each other inside the algebra $\mathfrak{sp}\big(2(d+3)\big)$.

In the following, we will be interested in the Carrollian limits (electric and magnetic) of the previous description and their realisation in the bulk and on the boundary. 
The Carrollian contraction of AdS$_{d+2}$ spacetime, called here AdS-Carroll spacetime, is homeomorphic to the manifold $\mathbb R \times H_{d+1}$, where $\mathbb R$ spans the real line of Carrollian time and $H_{d+1}$ is the $(d+1)-$dimensional hyperbolic space. 
Choosing angular coordinates, the AdS-Carroll metric with unit\footnote{The AdS-Carroll curvature radius, which is the curvature radius of the hyperbolic plane $H_d$, plays little role and can be set to one. 
The conformal boundary of AdS-Carroll$_{d+2}\cong \mathbb R \times H_{d+1}$ is $\mathscr I_{d+1}\cong\mathbb{R}\times S^d$ and the celestial sphere $S^d$ also has unit curvature.} curvature radius reads
\begin{equation} \label{eq:AdS-Carroll-metric}
    {\rm d}s_\text{AdS-Carroll$_{d+2}$}^2 = {\rm d}\ell_{H_{d+1}}^2 = {\rm d} \xi^2 + \sinh^2\xi\,{\rm d}\Omega_d^2 \,,
\end{equation}
where $\mathbf x$ are angular coordinates on the $d-$dimensional sphere. The pull-back of the AdS-Carroll metric on constant $\xi$ hypersurfaces defines a metric which is related, up to a conformal factor of $\sinh^2 \xi$, to the metric of $\mathscr I_{d+1}$. The latter is seen as the boundary of AdS-Carroll and is located in the asymptotic region where $|u| \gg 1$ and $\xi \gg 1$ such that $u_\text{bdy} := u/\cosh\xi$ remains finite. This contrasts with the AdS case where the time-like coordinate remains finite and deserves some explanation. Recall that, in global coordinates, AdS$_{d+2}$ spacetime with unit curvature radius can be described by the locus $X^2 = -1$ of the set of coordinates $X^A$, $A \in \{0,1,\ldots,d+2\}$ in ambient space. 
A convenient parameterisation makes use of (hyperbolic) angular coordinates
\begin{subequations}
\begin{align}
    X^0 &= \alpha \cosh \xi\,\sin \tau \,, \\
    X^i &= \alpha \sinh\xi\,\hat x^i \,, \\
    X^{d+2} &= \alpha \cosh\xi\,\cos\tau \,,
\end{align}
\end{subequations}
where $\alpha \geqslant 0$, $\xi \geqslant 0$, $\tau \in [0,2\pi[$ and $\sum_{i=1}^{d+1} (\hat x^i)^2 = 1$. The angular coordinates $\mathbf x$ on the celestial sphere correspond to the usual parameterisation of the $d-$dimensional sphere with unit curvature embedded in the $(d+1)-$dimensional Euclidean space with coordinates $\hat x^i$ in terms of $d$ angles. 
With this parameterisation, the Euler operator reads $X^A \partial_A$. The bulk of AdS with unit curvature radius is then obtained by setting $\alpha = 1$.

Redefining the time coordinate by $\tau = c\,u_\text{bdy}$, the Carrollian limit $c \to 0$ of the previous parameterisation gives
\begin{subequations} \label{eq:AdS-Carroll-ambient}
\begin{align}
    u &:= \lim_{c \to 0} c^{-1} X^0 = \alpha \cosh \xi\,u_\text{bdy} \,, \\
    y^i &:= \lim_{c \to 0} X^i = \alpha \sinh\xi\,\hat x^i \,,\\
    y^{d+2} &:= \lim_{c \to 0} X^{d+2} = \alpha \cosh\xi \,,
\end{align}
\end{subequations}
where now $u_\text{bdy} \in \mathbb R$.\footnote{This can be seen by noticing that there is no constraint on $u_\text{bulk}$ after the limit is taken. 
Alternatively, one can switch to the universal cover of AdS where $\tau \in \mathbb R$ before sending $c \to 0$.} The coordinate $u$ differs from boundary time $u_\text{bdy}$ by a conformal factor $\cosh\xi$, which diverges when $\xi \to \infty$. 
The bulk of AdS-Carroll is still obtained by setting $\alpha = 1$. 
Note however that the Euler operator $u \partial_u + y^a \partial_a$ on the ambient space description of AdS-Carroll reads $u \partial_u + \alpha \partial_\alpha$.

%%%%%%%%%%%%%%%%%%
\subsection{Electric} \label{sec:ambient-electric}

As explained in the introductory remarks of this section, the ambient description of the electric theory can be obtained as the $c \to 0$ limit of the ambient description of the AdS singleton, obtained by performing the splitting $X^A = (c u,y^a)$ where \mbox{$a\in\{1,2,\ldots,d+2\}$}. 
After rescaling the first of \eqref{eq:ambient-relativistic} by $-c^2$, we obtain
\begin{equation} \label{eq:ambient-electric}
    \partial_u{}^2\,\Phi = 0 \,,\quad \left(u\,\partial_u + y^a \partial_a + \Delta_-\right) \Phi = 0 \,,\quad \Phi \simeq \Phi + y^2\,\Psi \,,
\end{equation}
for $\Delta_- = \frac{d-1}{2}$ and any $\Psi$ satisfying $\left(u\,\partial_u + y^a \partial_a + \Delta_- + 2\right) \Psi = 0$ and $\partial_u{}^2 \Psi = 0$ \cite{Bekaert:2022oeh}. 
The value of $\Delta_-$ is inherited from the relativistic parent but comes as an extra piece of information from a purely Carrollian perspective, since the scaling dimension is not fixed anymore by the requirement that the constraints close on a Lie algebra. Instead, as was remarked in \cite{Bekaert:2022oeh}, these constraints always close on a contraction of $\mathfrak{sp}(2,\mathbb{R})\cong \mathfrak{so}(2,1)$ isomorphic to $\mathfrak{iso}(1,1)$.

The isometries of the Carrollian-like ambient space can be seen as the maximal set of vector fields that commute with the constraints \eqref{eq:ambient-electric} and are given by
\be \label{eq:ambient-isometries-electric}
J_{ab} = 2 y_{[a} \partial_{b]} \,,\quad P_a = y_a \partial_u \,,
\ee
acting as Lorentz transformations of constant-$u$  hyperplanes $\mathbb R^{d+1,1}\subset \mathbb R^{d+1,2}$ and `Carroll' boosts in the $u$ direction. 
These isometries are also the Carrollian contraction of the algebra $\mathfrak{so}(d+1,2)$ to $\mathfrak{iso}(d+1,1)$. 
In other words, the constraint algebra $\mathfrak{iso}(1,1)$ and the isometry $\mathfrak{iso}(d+1,1)$ form a Howe dual pair in $\mathfrak{sp}\left(2(d+3)\right)$.

A bulk AdS-Carroll description for the electric simpleton, inspired by the case of the AdS singleton, is given by evaluating the ambient field $\Phi$ verifying Eq.~\eqref{eq:ambient-electric} on the submanifold $y^2 = -1$, which is a hyperbolic cylinder in ambient space: $\mathbb{R}\times H_{d+1}\subset\mathbb R^{d+1,2}$. 
The pullback  $\upvarphi=\Phi|_{y^2=-1}$ is a field on AdS-Carroll spacetime with equation of motion
\be \label{eq:AdSC-electric}
    \partial_u{}^2 \upvarphi(u,\xi,\mathbf{x}) = 0 \,,
\ee
which is a straightforward consequence of the first of \eqref{eq:ambient-electric}. 
Compared with the AdS case, it is not so easy to see how the other two conditions can be implemented.

Indeed, in AdS, there are generically two branches of solutions to the Klein-Gordon equation for a scalar field. 
The quotienting condition in ambient space (i.e. the last of Eq.~\eqref{eq:ambient-relativistic}) is a consequence of the fine-tuning of the mass (or scaling dimension), and can be understood as quotienting by the subleading branch of solutions corresponding to a scaling in the holographic coordinate with power equal to the biggest solution in $\Delta$ to the mass-shell equation. 
The quotienting condition is therefore what distinguishes the singleton from a generic scalar field, see e.g. \cite{Bekaert:2013zya}.

In AdS-Carroll, the Klein-Gordon equation reduces to Eq.~\eqref{eq:AdSC-electric} in the electric limit. Therefore, no mass-shell condition exists, nor are there two distinct branches of solutions. 
Nevertheless, one can implement the quotienting condition in ambient space and constrain the form of the solutions of Eq.~\eqref{eq:AdSC-electric}. 
Using the analysis of \cite{Bekaert:2013zya}, in the coordinates of Eq.~\eqref{eq:AdS-Carroll-ambient} gives the following expression for the simplest representative in ambient space
\begin{equation}
    \Phi(\alpha,u,\xi,\mathbf x) = \left(\alpha \sinh\xi\right)^{-\Delta_-} \varphi_-(\mathbf x) + u \left(\alpha \sinh\xi\right)^{-\Delta_+} \varphi_+(\mathbf x) \,,
\end{equation}
where the $u$, $\alpha$ and $\xi$ dependence are a consequence of the equation of motion, the homogeneity condition and the quotienting condition respectively. From the scaling in $\alpha$, it can be seen that $\varphi_\pm$ have boundary scaling dimension $\Delta_\pm$, in accordance with the analysis of Section~\ref{electric/magnetic}. 
While the linearity in $u$ and the scaling in $\alpha$ can be inferred directly from the equation of motion and the homogeneity condition, the $\xi$ dependence can only be revealed by an analysis near the conformal boundary of AdS-Carroll, where $\xi \gg 1$.
The bulk field $\upvarphi$ can then be obtained by evaluating $\Phi(\alpha,u,\xi,\mathbf x)$ at $\alpha = 1$ and yields
\begin{equation} \label{eq:bulk-electric}
    \upvarphi(u,\xi,\mathbf x) = \frac{1}{(\sinh\xi)^{\Delta_-}} \left(\varphi_-(\mathbf x) + \frac{u}{\sinh\xi}\varphi_+(\mathbf x) \right) .
\end{equation}

The boundary description of the simpleton \eqref{eq:boundary-electric} was obtained in \cite{Bekaert:2022oeh} by performing the null projection of the defining equations \eqref{eq:ambient-electric} $(u,y^a) \sim \lambda\,(u,y^a)$ for all $\lambda > 0$ along the light-cone $y^2 = 0$.
In the context of a bulk-boundary correspondence in AdS-Carroll, we can now also access it by focusing on the boundary dynamics for the field $\upvarphi(u,\xi,\mathbf x)$ defined in Eq.~\eqref{eq:bulk-electric}, obtained for $\xi \gg 1$ and identifying the boundary time using $u = u_\text{bdy}\,\cosh \xi$
\begin{equation}
    \varphi(u_\text{bdy},\mathbf x) := \lim_{\xi \to \infty} \big[\,(\sinh\xi)^{\Delta_-} \upvarphi(u,\xi,\mathbf x)\,\big] = \varphi_-(\mathbf x) + u_\text{bdy}\,\varphi_+(\mathbf x) \,.
\end{equation}
We find that the simpleton is described on the boundary by the electric description in Section~\ref{electric/magnetic}, i.e.\ by a Carrollian primary field $\varphi(u_\text{bdy},\mathbf x)$ of scaling dimension $\Delta_- = \frac{d-1}{2}$ verifying Eq.~\eqref{eq:boundary-electric-eom} and thus parameterised by two arbitrary functions of the angles $\varphi_\pm(\mathbf x)$ of scaling dimensions $\Delta_\pm$ as in
\eqref{gensol}. 

%%%%%%%%%%%%%%%%%%
\subsection{Magnetic} \label{sec:ambient-magnetic}

The magnetic limit is obtained by switching to first-order (`Hamiltonian') variables before sending $c$ to zero. Going back to Eq.~\eqref{eq:ambient-relativistic} and defining $\Pi = \frac{1}{c^2} \partial_u \Phi$, we obtain the following system of constraints in the limit $c \to 0$
\begin{subequations} \label{eq:ambient-magnetic}
\begin{align}
    \partial_u \Phi &= 0 \,,&\quad \left(y^a \partial_a + \Delta_-\right) \Phi &= 0 \,,&\quad \Phi &\simeq \Phi + y^2 f \,,\\
    \partial_u \Pi &= \Box_{{}_{\mathbb R^{d+1,1}}} \Phi \,,&\quad \left(u\,\partial_u + y^a \partial_a + \Delta_+\right) \Pi &= 0 \,,&\quad \Pi &\simeq \Pi - 2u f + y^2 g \,,
\end{align}
\end{subequations}
where, as usual, $\Delta_\pm = \frac{d \pm 1}{2}$ and we already dropped the $u\,\partial_u$ term in the homogeneity constraint for $\Phi$ because of the equation of motion $\partial_u \Phi = 0\,$. 
The functions $f$ and $g$ verify the same equations of motion as $\Phi$ and $\Pi$ respectively (i.e. $\partial_u f = 0$ and $\partial_u g = \Box_{{}_{\mathbb R^{d+1,1}}} f$) and are homogeneous of degree $-(\Delta_- + 2)$ and $-(\Delta_+ + 2)$ respectively. 
We can recast the previous set of equations into matrix form as follows
\begin{subequations}
\begin{align}
    \begin{pmatrix} \partial_u & 0 \\ - \Box_{{}_{\mathbb R^{d+1,1}}} & \partial_u \end{pmatrix} \begin{pmatrix} \Phi \\ \Pi \end{pmatrix} &= 0 \,,\\
    \begin{pmatrix} y^a \partial_a + \Delta_- & 0 \\ 0 & u\,\partial_u + y^a \partial_a + \Delta_+ \end{pmatrix} \begin{pmatrix} \Phi \\ \Pi \end{pmatrix} &= 0 \,,\\
    \begin{pmatrix} \Phi \\ \Pi \end{pmatrix} &\simeq \begin{pmatrix} \Phi \\ \Pi \end{pmatrix} + \begin{pmatrix} y^2 & 0 \\ - 2u & y^2 \end{pmatrix} \begin{pmatrix} f \\ g \end{pmatrix} .
\end{align}
\end{subequations}
The set of constraints spanned by the three $2 \times 2$ matrices form an algebra which closes on the same contraction of $\mathfrak{sp}(2,\mathbb{R})$ as that generated by the constraints \eqref{eq:ambient-electric}, i.e.\ $\mathfrak{iso}(1,1)$. 
Moreover, the set of matrix-valued vector fields commuting with these matrices (the maximal commuting pair) is given by
\begin{equation}
    J_{ab} =
    \begin{pmatrix}
        2\,y_{[a} \partial_{b]} & 0 \\ 0 & 2\,y_{[a} \partial_{b]}
    \end{pmatrix}
    ,\quad
    P_a =
    \begin{pmatrix}
        y_a \partial_u & 0 \\ \partial_a & y_a \partial_u
    \end{pmatrix}
    ,
\end{equation}
generating again the $(d+2)$-dimensional Poincar\'e algebra.

The bulk description of the magnetic theory in AdS-Carroll is built using a doublet of scalar fields in $\mathbb R \times H_{d+1}$ verifying
\begin{equation} \label{eq:bulk-magnetic}
    \partial_u \upphi = 0 \,,\quad \partial_u \uppi = \left(\nabla_{H_{d+1}}^2 - \Delta_- (\Delta_- - d)\right) \upphi = \left(\nabla_{H_{d+1}}^2 + \tfrac{d^2-1}{4}\right) \upphi \,.
\end{equation}
Applying the same techniques as in the electric case, we can find a representative which, in the coordinates of Eq.~\eqref{eq:AdS-Carroll-metric}, reads
\begin{equation} \label{eq:magnetic-AdS-Carroll}
    \upphi(u,\xi,\mathbf x) = \frac{1}{(\sinh\xi)^{\Delta_-}} \psi_-(\mathbf x) \,,\quad \uppi(u,\xi,\mathbf x) = \frac{1}{(\sinh\xi)^{\Delta_+}} \left(\psi_+(\mathbf x) + \frac{u}{\sinh \xi} \hat\nabla^2 \psi_-(\mathbf x) \right) ,
\end{equation}
where $u$ is the null time coordinate and it can be seen from the $\sinh\xi$ dependence that $\psi_\pm(\mathbf x)$ have scaling dimensions $\Delta_\pm$ respectively.

As usual, the boundary description of \eqref{eq:ambient-magnetic} is obtained by performing the null projection $(u,y^a) \sim \lambda\,(u,y^a)$ for all $\lambda >0$ along the light-cone $y^2 = 0$. 
Alternatively, it can be directly read off from Eq.~\eqref{eq:magnetic-AdS-Carroll} and matched with Eq.~\eqref{eq:magnetic-on-shell} by identifying again $u = u_\text{bdy}\,\cosh \xi$ and considering
\begin{subequations}
\begin{align}
    \phi(u_\text{bdy},\mathbf x) &:= \lim_{\xi \to \infty} \big[\,(\sinh\xi)^{\Delta_-}\xi\ \upphi(u,\xi,\mathbf x)\,\big] = \psi_-(\mathbf x) \,,\\
    \pi(u_\text{bdy},\mathbf x) &:= \lim_{\xi \to \infty} \big[\,(\sinh\xi)^{\Delta_+}\ \uppi(u,\xi,\mathbf x)\,] = \psi_+(\mathbf x) + u_\text{bdy} \hat\nabla^2\psi_-(\mathbf x) \,,
\end{align}
\end{subequations}
which is indeed the on-shell description of the magnetic theory \eqref{eq:magnetic-on-shell}.

%%%%%%%%%%%%%%%%%%%%%%%%%%%%%%%%%%%%
\section{Discussion and outlook}\label{sec:conclusions}

In this paper, we discussed the non-unitary indecomposable\footnote{Indecomposable representations already appeared in the context of flat holography; see, e.g., \cite{Bagchi:2012yk, Fiorucci:2023lpb}.} representation of the Carroll, Poincar\'e and BMS algebras that is realised on the space of solutions of a conformal Carrollian scalar defined on $\mathscr{I}_{d+1} \cong \mathbb R \times S^d\,$. 
We dubbed it as \emph{simpleton} and we showed that it can be recovered as a smooth limit of Dirac's singleton representation of the conformal algebra $\mathfrak{so}(d+1,2)$. 
We also showed that corresponding smooth limits can be introduced for various realisations of Dirac's singleton. This allowed us to realise the simpleton holographically, in terms of bulk field theories defined either on $(d+2)-$dimensional Minkowski or on AdS-Carroll spacetimes.
We achieved these results either directly or by embedding the Minkowski and AdS-Carroll spacetimes in an ambient space with yet another extra dimension.

At this stage, these holographic realisations only amount to an identification of the space of solutions of the various field theories involved. 
Still, these should be considered as the first steps in identifying a flat analogue of higher-spin holography, as motivated by the role played by the singleton representation in this context. 
A natural next step towards the realisation of the proposed flat limit of higher-spin holography would be to study the limit of the Flato-Fronsdal theorem \cite{Flato:1978qz}, which is considered as a pillar of such a duality. 
In brief, this theorem states that the tensor product of two singletons decomposes into an infinite sum of massless higher-spin fields in AdS. 
This has been considered as a prediction for the spectrum of the boundary dual of a relativistic conformal scalar living on the boundary of AdS, while it remains unclear how the product of two simpletons could account for the degrees of freedom of propagating fields in Minkowski space. 
Local degrees of freedom might be encoded in a much subtler way in Carrollian holography, so that a further study of this aspect might teach interesting lessons for flat holography. 

Let us also point out that higher-spin holography could also admit an ultra-relativistic limit on both sides, involving, e.g., a Carrollian limit of Vasiliev's theory on a bulk AdS-Carroll spacetime. 
While looking rather exotic, this option could indirectly provide useful hints about flat holography. 
For instance, the AdS-Carroll spacetime $\mathbb R \times H_{d+1}$ is the blowup of time-like infinity for Minkowski spacetime $\mathbb R^{d+1,1}$ \cite{Figueroa-OFarrill:2021sxz, Have:2024dff}.

Both electric and magnetic theories can be coupled to a more generic class of background metrics \cite{Rivera-Betancour:2022lkc, Baiguera:2022lsw}.\footnote{In particular, the magnetic theory can be coupled to shearless Carrollian metrics without losing invariance under Carroll boosts and these classes of metrics naturally emerge at null infinity while considering the limit of Einstein spaces; see, e.g., \cite{Campoleoni:2023fug}.}
We only analysed the case of $\mathscr{I}_{d+1} \cong \mathbb R \times S^d\,$, but it would be interesting to perform a more systematic analysis, coupling to the most general Carrollian metric (which could be interpreted as an external source in the Carrollian CFT and thus play a role in a putative holographic setup). 
In particular, it would be interesting to check whether the equivalence between electric and magnetic theories also holds in more general backgrounds. Along similar lines, it would be important to also test the robustness of our findings against the introduction of self-interactions for the Carrollian scalar. Encouraging results were obtained in \cite{Barnich:2012rz, Barnich:2014xnb}, where a BMS$_3-$invariant field theory given by a magnetic Carrollian scalar with potential was shown to control the boundary dynamics at null infinity of three-dimensional asymptotically-flat gravity in the classical regime. 
In \cite{Barnich:2012rz} a non-local map between the `time-like' and the `space-like' theories, analogous to that discussed in Section~\ref{electric/magnetic}, was also introduced. 
Polynomial interactions without derivatives (similar to those playing a crucial role in higher-spin holography \cite{Giombi:2016ejx}) were also studied for Carrollian scalars in \cite{Banerjee:2023jpi}. 
Quartic interactions involving derivatives have also been considered in higher-spin holography \cite{Giombi:2013yva} via double-trace deformations which are the square of conformal currents. 
Similar interactions have been already considered in the fracton model that was proposed in \cite{Pretko:2018jbi} (see also \cite{Bidussi:2021nmp}). 
When seen as a Carrollian field theory, the latter involves the electric action \eqref{eq:boundary-electric} as kinetic term, to which one adds a quartic interaction with four derivatives of the `double-trace' type $\mathcal{O}\sim\tfrac1{N} J_{ij}J^{ij}$, where $J_{ij}\sim \phi\stackrel{\leftrightarrow}{\partial}_i\stackrel{\leftrightarrow}{\partial}_j\phi$ stands for the bilinear spin-two current (or stress-tensor). Dimensional analysis shows that, near the Gaussian fixed point, the scaling dimension of the bilinear tensor is $\Delta^{free}_{J_2}=d+1$. Furthermore, the usual Hubbard-Stratonovitch trick applied to double-trace deformations leads to an anomalous dimension $\Delta^{int}_{J_2}=d+1-\Delta^{free}_{J_2}=0$ at the interacting fixed point in the large-$N$ limit.
In AdS/CFT context, this situation is the one of holographic degeneracy where the two fixed point are described by the same bulk theory with Dirichlet versus Neumann boundary conditions. 
By analogy with higher-spin holography, it is tempting to speculate that the large-$N$ limit of this fracton model is dual to a putative higher-spin theory around Minkowski spacetime where the free and interacting fixed points of the Carrollian conformal field theory are described by considering two distinct asymptotic behaviours for the gravitational field. 
Such a scenario might also provide a fracton gravity analogue of the holographic description of induced higher-spin gravity \cite{Giombi:2013yva}.

Finally, it would also be interesting to investigate the flat/Carrollian limit of the higher-order singletons considered in \cite{Bekaert:2013zya}. 
To this end, one can follow the same lines as in Section~\ref{sec:limit-field}, but replacing the singleton scaling dimension with the more general values $\Delta=\frac{d+1-2\ell}2$ for $\ell\in\mathbb N$. 
From the conformal boundary perspective, the ultra-relativistic limit of higher-order singletons described by the polywave equation $\Box^\ell\phi=0$ corresponds to higher-order (electric) simpletons obeying to the equation $\partial_u^{2\ell}\varphi=0$.
From a group-theoretical perspective, they are described by quotients 
\be
\mathcal \cW^{\mathfrak{iso}(d+1,1)}\left(\tfrac{d+1-2\ell}2,0\right) =
\mathcal V^{\mathfrak{iso}(d+1,1)}\left(\tfrac{d+1-2\ell}2,0\right) \,\big /\, \mathcal V^{\mathfrak{iso}(d+1,1)}\left(\tfrac{d+1+2\ell}2,0\right) .
\ee
Their first-order (magnetic) description should arise from the asymptotic expansion of massless scalar modes with power of $1/r$ in the range $\frac{d+1}2-\ell$, $\frac{d+1}2-\ell+1$, $\dots$, $\frac{d-1}2+\ell-1$, $\frac{d-1}2+\ell$, in agreement with the branching rule \eqref{orderNquotient} for the restriction $\mathfrak{iso}(d+1,1)\downarrow\mathfrak{so}(d+1,1)$ by setting $\Delta=\tfrac{d+1-2\ell}2$ and $N=2\ell$.
Furthermore, these scaling dimensions also agree with the branching rule of the higher-order singleton for the restriction $\mathfrak{so}(d+1,2)\downarrow\mathfrak{so}(d+1,1)$.

\section*{Acknowledgments}

We thank G.~Barnich, T.~Basile, N.~Boulanger, D.~Grumiller, S.~Prohazka, S.~I.~A.~Raj and M.~Vilatte for useful comments and discussions.
We thank the \'Ecole polytechnique (X.B. and A.C.), the University of Mons (X.B. and S.P.), the University of Edinburgh (S.P.) and the University of Tours (S.P.) for hospitality.
The work of A.C.\ and S.P.\ was partially supported by the Fonds de la Recherche Scientifique -- FNRS under Grants No.\ FC.36447, F.4503.20, T.0022.19 and T.0047.24. The work of S.P.\ is funded by the \textit{Fonds Friedmann} run by the \textit{Fondation de l'\'Ecole polytechnique}.
This research was partially completed during the workshop `Fluids, black holes, conformal systems and null infinity' at \'Ecole polytechnique, the `5${}^\text{th}$ Mons Workshop on Higher Spin Gauge Theories' at the University of Mons, and the thematic programme `Carrollian Physics and Holography' at the Erwin Schr\"odinger Institute of Vienna, where preliminary versions of these results were also presented. 

%%%%%%%%%%%%%%%%%%%%%%%%%%%%%%%%%%%%%%%%%%%%%%%%%%%%%%%%

\appendix

%%%%%%%%%%%%%%%%%%%%%%%%%%%%%%%%%%%%
\section{Flat/Carroll limit of the bulk/boundary symmetries} \label{sec:limits-bulk-bdy}

The Poincar\'e algebra $\mathfrak{iso}(d+1,1)$ is the flat ($R \to \infty$) limit of the AdS isometry algebra $\mathfrak{so}(d+1,2)$, but it is also isomorphic to the Carrollian ($c \to 0$) contraction of the conformal algebra $\mathfrak{so}(d+1,2)$.
In fact, the inverse of the curvature radius may be identified with the velocity of light ($c=1/R$) on the AdS boundary in the flat limit AdS$_{d+2}\stackrel{R\to\infty}{\longrightarrow}\mathbb{R}^{d+1,1}$. 
The Poincar\'e algebra is also obtained as the Carrollian contraction of the AdS isometry algebra in the Carrollian limit from AdS spacetime to AdS-Carroll spacetime. 

In order to recall how to see these facts, consider the generators $\mathbb J_{AB}$ of $\mathfrak{so}(d+1,2)$ with Lie brackets
\be
\left[\mathbb J_{AB}, \mathbb J_{CD}\right] = \eta_{BC}\,\mathbb J_{AD} - \eta_{AC}\,\mathbb J_{BD} - \eta_{BD}\,\mathbb J_{AC} + \eta_{AD}\,\mathbb J_{BC}
\ee
where capital Latin indices $A,B,C,D \in \{0,1,\,\dots,\, d+1,0'\}$ and
the ambient metric $\eta_{AB}$ has signature $(-,+,\dots,+,-)$.
Picking a time-like direction, say $0'$, considered as the unphysical time direction, one can split and rescale the generators $\mathbb J_{AB}$ into the generators $\mathbb P_{a}:=\frac1{R}\,\mathbb J_{a0'}$ of transvections and the generators $\mathbb J_{ab}$ of the Lorentz subalgebra $\mathfrak{so}(d+1,1)$, with $a,b,c,d \in \{0,1,\,\dots,\, d+1\}$.
The Lie brackets of the AdS isometry algebra in $d+2$ dimensions read
\be \label{eq:AdS-conformal}
\begin{split}
\left[\mathbb J_{ab}, \mathbb J_{cd}\right] &= \eta_{bc}\,\mathbb J_{ad} - \eta_{ac}\,\mathbb J_{bd} - \eta_{bd}\,\mathbb J_{ac} + \eta_{ad}\,\mathbb J_{bc} \,,\\
\left[\mathbb J_{ab}, \mathbb P_c\right] &= \eta_{bc}\,\mathbb P_a - \eta_{ac}\,\mathbb P_b \,,\\
\left[\mathbb P_a, \mathbb P_b\right] &= \frac1{R^2}\, \mathbb J_{ab} \,,
\end{split}
\ee
where the Minkowski metric $\eta_{ab}$ has signature $(-,+,\dots,+)$ and, to distinguish more easily between the various realisations of the conformal algebra, in this appendix we denote generators in $d+2$ dimensions by blackboard bold letters (differently from the main body of the text). The Poincar\'e algebra \mbox{$\mathfrak{iso}(d+1,1)$} is obviously the flat ($R \to \infty$) limit of the AdS isometry algebra $\mathfrak{so}(d+1,2)$.
Picking the other time-like direction $0$, considered as the physical time direction, and taking a similar limit with the substitution $R=1/c$ would of course provide an isomorphic algebra. However, the latter would be interpreted physically as an ultra-relativistic limit ($c\to 0$) of the AdS isometry algebra.\footnote{Note that in the latter contraction $c\to 0$, the Lorentz subalgebra $\mathfrak{so}(d+1,1)\subset\mathfrak{so}(d+1,2)$ is contracted to the homogeneous Carroll algebra $\mathfrak{iso}(d+1)\subset \mathfrak{iso}(d+1,1)$ while in the former contraction the Lorentz subalgebra is unaffected. This explains why the corresponding homogeneous spaces (Minkowski vs AdS-Carroll) are distinct although they have isomorphic isometry algebras (see, e.g., \cite{Figueroa-OFarrill:2018ilb}).}

Consider now the conformal algebra $\mathfrak{so}(d+1,2)$ generated by Lorentz transformations $J_{\mu\nu}\,$, translations $P_\mu\,$, special conformal transformations $K_\mu$ and dilations $D$ and satisfying the Lie brackets
\be
\begin{split} \label{eq:conformal-algebra}
[J_{\mu\nu}, J_{\rho\sigma}] &= \eta_{\nu\rho}\,J_{\mu\sigma} - \eta_{\mu\rho}\,J_{\nu\sigma} - \eta_{\nu\sigma}\,J_{\mu\rho} + \eta_{\mu\sigma}\,J_{\nu\rho} \,,\\
[J_{\mu\nu}, P_\rho] &= \eta_{\nu\rho}\,P_\mu - \eta_{\mu\rho}\,P_\nu \,,\quad [D, P_\mu] = P_\mu \,,\\
[J_{\mu\nu}, K_\rho] &= \eta_{\nu\rho}\,K_\mu - \eta_{\mu\rho}\,K_\nu \,,\quad [D, K_\mu] = -K_\mu \,,\\
[K_\mu, P_\nu] &= 2\,\eta_{\mu\nu}\,D - 2\,J_{\mu\nu} \,,
\end{split}
\ee
with $\mu,\nu,\rho,\sigma \in \{0,\,\dots,\, d \}$ and where the metric $\eta_{\mu\nu}$ has signature $(-,+,\dots,+)\,$. 
Splitting the components between space ($\mu = i$) and time ($\mu = 0$), and rescaling the generators with a time-like component, as $B_i = c\,J_{i0}\,$, $H = c\,P_0$ and $K = c\,K_0$, the conformal algebra can be seen to generate the isometries of AdS spacetime. 
Let us define the generators
\be
\mathbb P_0 = \tfrac12 (H - K) \,,\quad \mathbb P_i = - B_i \,,\quad \mathbb P_{d+1} = \tfrac12 (H + K) \,,
\ee
and
\be
\mathbb J_{i0} = \tfrac12 (P_i - K_i) \,,\quad \mathbb J_{ij} = J_{ij} \,,\quad \mathbb J_{i\,d+1} = \tfrac12 (P_i + K_i) \,,\quad \mathbb J_{0\,d+1} = D \,.
\ee
One may check that they verify the Lie brackets of the AdS isometry algebra \eqref{eq:AdS-conformal} in $d+2$ dimensions upon setting $c=1/R$.

%%%%%%%%%%%%%%%%%%%%%%%%%%%%%%%%%%%%
\section{Electric-magnetic duality in Carrollian theories} \label{sec:duality}

The electric and magnetic limits of relativistic theories that lead to Carrollian theories were first discussed on-shell for electromagnetism in \cite{Duval:2014uoa} and later generalised off-shell and for massless fields of any spin in \cite{Henneaux:2021yzg}.
In general, these limits lead to inequivalent theories (as emphasised in \cite{Henneaux:2021yzg}) but for the case of spin-$s$ simpletons (i.e.\ for the ultra-relativistic limit of spin-$s$ singletons\footnote{In this appendix, we denote as singletons all representations of the conformal algebra that remain irreducible (or split at most in two) upon restriction to its Poincar\'e subalgebra. In a spacetime of dimension four, they correspond to massless representations of any spin \cite{Fronsdal:1978vb, Angelopoulos:1980wg, Bekaert:2011js}.}), they are related by an electric-magnetic duality transformation. In particular, this is true for all Carrollian spin-$s$ field with zero energy in spacetime dimension four.

For the $s=1$ example, the on-shell equivalence of the electric and magnetic limits of Maxwell equations was pointed out at the end of Section~V in \cite{Duval:2014uoa}. 
The off-shell equivalence of electric and magnetic formulations of the spin-one simpleton (Carrollian electromagnetism) in spacetime dimension $d+1=4$ is manifest if one solves the Gauss constraint $\partial_a\pi^a=0$ explicitly as $\pi^a=\epsilon^{abc}\partial_b Z_c$ ($a,b,c=1,2,3$) inside the electric and magnetic actions
\be\label{eq:ElectricAction} 
S_{\text{el}}[A_a, \pi^a, A_u] = \int {\rm d}u \,{\rm d}^d x \left(\pi^a \dot{A}_a -\frac12 \pi^a \pi_a+ A_u \partial_a \pi^a \right) ,  
\ee
and
\be\label{eq:MagneticAction} 
S_{\text{mag}}[A_a, \pi^a, A_u] = \int {\rm d}u\, {\rm d}^d x \left(\pi^a \dot{A}_a -\frac14 F^{ab} F_{ab}+ A_u \partial_a \pi^a \right) , 
\ee
corresponding to the equations
(5.13) and (5.18) of \cite{Henneaux:2021yzg}. The resulting actions are manifestly mapped into each other upon exchanges $A_a \leftrightarrow Z_a$. 
Equivalently, the corresponding actions can also be obtained, respectively, from the electric and magnetic limits of the manifestly duality-invariant formulation of Maxwell electromagnetism \cite{Deser:1976iy}. 
In order to be more precise about this equivalence, remember that an electric-magnetic duality transformation is local in terms of field strengths but non-local in terms of potentials (however, it is non-local in space, but local in time).

For Carrollian tensor fields of spin $s\geqslant 2$ and zero energy in dimension four, one could analogously start from the action principles \cite{Henneaux:2004jw, Henneaux:2016zlu} where the constraints have been solved inside the action in terms of prepotentials. 
This generalises to the on-shell description of spin-$s$ singletons in higher dimensions since they are always invariant under electric-magnetic duality (see e.g. \cite{Fronsdal:1978vb, Angelopoulos:1980wg, Bekaert:2011js} for the precise definitions and properties of these generalisations). 

%%%%%%%%%%%%%%%%%%%%%%%%%%%%%%%%%%%%
\section{Higher symmetries of the magnetic description} \label{sec:symmetries-magnetic}

In this appendix, we analyse the higher symmetries of the Carrollian theory \eqref{eq:boundary-magnetic}, that is the global symmetries that are realised in terms of higher-order differential operators acting on the fields. The higher symmetries of the electric theory \eqref{eq:boundary-electric} were given in \cite{Bekaert:2022oeh}.

One can conveniently rewrite the action \eqref{eq:boundary-magnetic} in a more compact form 
\be \label{eq:magnetic}
S_\text{mag}[\boldsymbol{\phi}] = \frac12 \int {\rm d}u \,{\rm d}^d \mathbf x\ \sqrt{\gamma}\ \boldsymbol{\phi}^\dagger\, \boldsymbol{K} \, \boldsymbol{\phi} \,,
\ee
with a doublet structure
\be\label{operatorK}
\boldsymbol{\phi} = \begin{pmatrix} \phi \\ \pi \end{pmatrix} , \quad \boldsymbol{K} = \begin{pmatrix} \hat \nabla^2 & -\partial_u \\ \partial_u & 0 \end{pmatrix} .
\ee
This formulation of the magnetic conformal scalar is such that the kinetic operator $\boldsymbol K$ is manifestly Hermitian under the product $\langle \boldsymbol{\phi} | \boldsymbol{\psi} \rangle = \int {\rm d}u\, {\rm d}^d \mathbf x\ \sqrt{\gamma}\ \boldsymbol{\phi}^\dagger\, \boldsymbol{\psi}$.

The higher symmetries of the magnetic theory, in the doublet formulation of Eq.~\eqref{eq:magnetic} are, by definition, differential operators
\be\label{operatorD}
\boldsymbol D = \begin{pmatrix} M & N \\ P & Q \end{pmatrix} ,
\ee
that commute weakly with the kinetic operator $\boldsymbol K$, defined in \eqref{operatorK}, in the sense that
\begin{equation}
    \boldsymbol K \circ \boldsymbol D = \boldsymbol D^\dagger \circ \boldsymbol K \quad\Leftrightarrow\quad \left \lbrace \begin{aligned} \hat\nabla^2 \circ M - \partial_u \circ P &= M^\dagger \circ \hat \nabla^2 + P^\dagger \circ \partial_u \\ \hat \nabla^2 \circ N - \partial_u \circ Q &= - M^\dagger \circ \partial_u \\ \partial_u \circ M &= N^\dagger \circ \hat\nabla^2 + Q^\dagger \circ \partial_u \\ \partial_u \circ N &= - N^\dagger \circ \partial_u \end{aligned} \right. ,
\end{equation}
quotienting by trivial symmetries of the form
\begin{equation}
    \boldsymbol D=\boldsymbol D' \circ \boldsymbol K = \begin{pmatrix} M' \circ \hat \nabla^2 + N' \circ \partial_u & -M' \circ \partial_u \\ P' \circ \hat \nabla^2 + Q'\circ \partial_u & - P'\circ \partial_u \end{pmatrix} ,
\end{equation}
with $M'^\dagger = M'$, $Q'^\dagger = Q'$ and $N'^\dagger = P'$. 

Using the equivalence relation, we can look (without loss of generality) for representatives $M$, $N$, $P$ and $Q$ which are independent of $\partial_u$ (the price to pay is to allow for operators proportional to $\hat\nabla^2$ in $M$ and $P$).\footnote{We may also choose a representative of $\boldsymbol D$ including powers of $\partial_u$, but the first choice will prove to be more convenient to compute symmetries.} With this choice, the equations are equivalent to
\begin{subequations}
\begin{align}
\dot N &= 0 \,,&\quad N^\dagger &= - N \,,\\
\dot Q &= \hat\nabla^2 \circ N \,,&\quad Q^\dagger  &=M \,,\\
\dot P &= \hat\nabla^2 \circ M - M^\dagger \circ \hat\nabla^2 \,,&\quad P^\dagger &= - P \,,
\end{align}
\end{subequations}
and, in the end, the differential operators $M$, $N$, $P$ and $Q$ read
\begin{subequations}\label{MNPQ}
\begin{align}
M &= L_{-1} + i\,L_{+1} - i\,u\,K_{+1} \circ \hat\nabla^2 \,,\\
N &= i\,K_{+1} \,,\\
P &= i\,K_{-1} + u \left [ \hat\nabla^2,L_{-1} \right ] + i\,u\left\{ \hat\nabla^2, L_{+1} \right\} - i\,u^2\,\hat\nabla^2\circ K_{+1} \circ \hat\nabla^2 \,,\\
Q &= L_{-1} - i\,L_{+1} + i\,u\,\hat\nabla^2\circ K_{+1} \,,
\end{align}
\end{subequations}
and are parameterised by four arbitrary Hermitian operators on the celestial sphere, denoted here by $K_{\pm 1}$ and $L_{\pm 1}$, as was the case for the electric theory \cite{Bekaert:2022oeh}. The square (respectively, round) bracket stands for the (anti)commutator.

Moreover, the commutation relations of such operators satisfy the infinite-dimensional Lie algebra \mbox{$\mathcal H(S^d) \,\otimes\, \mathfrak{gl}(2,\mathbb R)$}, which was already identified as the algebra of (higher) symmetries of the electric formulation,\footnote{\label{equivalencesubtlety}This is not surprising since the space of solutions of the electric and magnetic scalars are equivalent on-shell. However, remember that the equivalence involves a non-local operation related to the inversion of the shifted Laplacian \eqref{shiftedLaplacian}. This subtlety explains why the equivalence of higher symmetries (which are local differential operators) must nevertheless be carefully verified.} which we display here in full. Consider a higher symmetry of the magnetic formulation,
\be \label{eq:full-symmetries-magnetic}
\boldsymbol{D} \left(L_{-1}, L_{+1}, K_{-1}, K_{+1}\right) ,
\ee
that is to say a differential operator \eqref{operatorD} where the entries are given by \eqref{MNPQ}.
The Lie bracket of such two differential operators (times the imaginary unit) verifies
\begin{equation}
    i\left[\boldsymbol{D}\left( L_{-1},L_{+1},K_{-1},K_{+1}\right) , \boldsymbol{D}\left( L'_{-1},L'_{+1},K'_{-1},K'_{+1}\right) \right] = \boldsymbol{D}\left( L''_{-1},L''_{+1},K''_{-1},K''_{+1}\right) ,
\end{equation}
where
\begin{subequations}
\begin{align}
L''_{-1} &= i[L_{-1},L'_{-1}] - i [L_{+1},L'_{+1}] - \frac{i}2 [K_{-1},K'_{+1}] - \frac{i}2 [K_{+1},K'_{-1}] \,,\\
L''_{+1} &= i[L_{-1},L'_{+1}] + i[L_{+1},L'_{-1}] + \frac12 \{K_{-1},K'_{+1}\} - \frac12 \{K_{+1},K'_{-1}\} \,,\\
K''_{-1} &= i[L_{-1},K'_{-1}] + i [K_{-1},L'_{-1}] + \{L_{+1},K'_{-1}\} - \{K_{-1},L'_{+1} \} \,,\\
K''_{+1} &= i[L_{-1},K'_{+1}] + i [K_{+1},L'_{-1}] - \{L_{+1}, K'_{+1}\} + \{K_{+1},L'_{+1}\} \,.
\end{align}
\end{subequations}
Among these higher symmetries of the magnetic simpleton sit differential operators of order one (or less) corresponding to more familiar symmetries which we now list.

\paragraph{Large $\mathfrak{u}(1)$ transformations:}

Symmetries of order zero are given by taking \mbox{$L_{-1} = \alpha(\mathbf x)$} for $\alpha \in \mathscr C^\infty(S^d,\mathbb R)$ and all the others zero. 
They generate large $\mathfrak{u}(1)$ transformations, which act non-diagonally
\be
\delta \phi = i \alpha \phi \,,\quad \delta \pi = i \alpha \pi + i u \left [ \hat\nabla^2, \alpha \right] \phi \,.
\ee

\paragraph{Generalised BMS symmetry:}

For order one symmetries corresponding to generalised BMS transformations as defined in \cite{Campiglia:2014yka}, we will take
\be
\begin{split}
    L_{-1} &= -i\,Y^j(\mathbf x)\nabla_j - \frac{i}{2} \nabla_j Y^j(\mathbf x) \,,\\
    L_{+1} &= \frac{1}{2d} \nabla_j Y^j(\mathbf x) \,,\\[5pt]
    K_{-1} &= -T(\mathbf x)\hat\nabla^2 - \nabla^j T(\mathbf x)\nabla_j \,,
\end{split}
\ee
with $T(\mathbf x)$ and $Y^j(\mathbf x)$ real arbitrary functions of $\mathbf x \in S^d\,$. 
It will prove more convenient to add a well-chosen trivial symmetry given by
\be
N' = - P' = -i (T + u\,\lambda) \,,
\ee
where $\lambda = \frac{1}{d} \nabla_i Y^i$, in order to get to
\begin{subequations}
\begin{align}
M &= -i\left\{ Y^j \partial_j + \lambda \left(u\,\partial_u + \Delta_-\right) + T\partial_u \right \} ,\\
N &= 0 \,,\\
P &= -i\left \{ u\left[\hat\nabla^2, Y^j\partial_j + \lambda \Delta_- \right] - 2u \lambda \hat\nabla^2 + \partial^j T \partial_j \right\} ,\\
Q &= -i\left \{ Y^j \partial_j + \lambda \left(u\,\partial_u + \Delta_+\right) + T\partial_u \right \} ,
\end{align}
\end{subequations}
such that the generalised BMS transformations take a more familiar form
\begin{align}
\delta \phi &= \left( Y^i \partial_i + \lambda (u \partial_u + \Delta_-) + T \partial_u \right) \phi \,,\\[10pt]
\delta \pi &= \left( Y^i \partial_i + \lambda (u \partial_u + \Delta_+) + T \partial_u \right) \pi \nonumber\\
&\quad+ \left( u\left[\hat\nabla^2, Y^i \partial_i + \lambda \Delta_- \right] - 2u \lambda \hat\nabla^2 + \partial^i T \partial_i \right) \phi \,.
\end{align}

\paragraph{Most general symmetries of order one:}

In addition to the generalised BMS symmetries presented above, there are two additional symmetries of order one parameterised by \mbox{$L_{+1} = W(\mathbf x)$} and $K_{+1} = Z(\mathbf x)$
\begin{subequations}
\begin{align}
M &= i\,W - i\,u\,Z \circ \hat\nabla^2 \,,\\
N &= i\,Z \,,\\
P &= i\,u\left\{ \hat\nabla^2, W \right\} - i\,u^2\,\hat\nabla^2\circ Z \circ \hat\nabla^2 \,,\\
Q &= - i\,W + i\,u\,\hat\nabla^2\circ Z \,.
\end{align}
\end{subequations}

All these symmetries of order one close on an algebra which is isomorphic to a generalisation of the Newman-Unti algebra at level $3$. 
As explained in \cite{Duval:2014lpa}, the Newman-Unti algebra at level $N$ is generated by vector fields $X$ verifying
\begin{equation}
    \mathscr L_X g = \lambda g \,,\quad \left(\mathscr L_\xi\right)^N X = 0 \,,
\end{equation}
where $g$ and $\xi$ are the Carrollian metric and the field of observers respectively. 
One can also define a `generalised' version of the Newman-Unti algebras where we relax the first condition (this amounts to allow $X$ to generate an arbitrary diffeomorphism of the $d-$dimensional sphere). 
Therefore, the Lie subalgebra of symmetries of order one is isomorphic to the $N=3$ instance of the generalised Newman-Unti algebra.

\paragraph{Electric-magnetic equivalence:} \label{sec:symmetries-duality}

We saw in Section~\ref{sec:boundary} that one can map the (boundary) magnetic action into the electric one by means of a non-local redefinition. 
In this appendix, we show (cf.\ the remark in Footnote~\ref{equivalencesubtlety}) that the same procedure leads to a matching of the symmetries spelled out in Eq.~\eqref{eq:full-symmetries-magnetic} with the symmetries of its electric counterpart.

Note that, taking into account Eqs.~\eqref{eq:non-local} and \eqref{eq:electric-from-magnetic}, the symmetries $M$, $N$, $P$ and $Q$ acting on \mbox{$\varphi = \frac{1}{\sqrt{-\hat\nabla^2}}\pi$} can be recast into
\be
\delta \varphi = \frac{i}{\sqrt{-\hat\nabla^2}}(\delta \pi) = i\,\tilde D\,\varphi \,,
\ee
with
\be
\tilde D = \tilde L_{-1} \circ id + 2\,\tilde L_{+1} \circ H_0 + \tilde K_{-1}\circ H_{-1} + \tilde K_{+1} \circ H_{+1} \,,
\ee
where $H_m$ were defined in \cite{Bekaert:2022oeh} and
\begin{subequations}
\begin{align}
\tilde L_{-1} &= \frac12 \frac{1}{\sqrt{-\hat\nabla^2}} \circ \left( -\left\{\hat\nabla^2, L_{-1} \right\} - i \left[\hat\nabla^2, L_{+1} \right] \right) \circ \frac{1}{\sqrt{-\hat\nabla^2}} \,,\\
\tilde L_{+1} &= \frac12 \frac{1}{\sqrt{-\hat\nabla^2}} \circ \left(- \left\{\hat\nabla^2, L_{+1} \right\} + i \left[\hat\nabla^2, L_{-1} \right] \right) \circ \frac{1}{\sqrt{-\hat\nabla^2}} \,,\\
\tilde K_{-1} &= - \frac{1}{\sqrt{-\hat\nabla^2}} \circ K_{-1} \circ \frac{1}{\sqrt{-\hat\nabla^2}} \,,\\
\tilde K_{+1} &= + \sqrt{-\hat\nabla^2} \circ K_{+1} \circ \sqrt{-\hat\nabla^2} \,.
\end{align}
\end{subequations}
Note that $\tilde L_{\pm 1}$ and $\tilde K_{\pm 1}$ are all Hermitian.

%%%%%%%%%%%%%%%%%%%%%%%%%%%%%%%%%%%%
\section{Doublet description from AdS}
%Bulk simpleton as a flat limit of the bulk singleton 
\label{sec:smooth-limit}

The goal of this appendix is to obtain a doublet formulation of a bulk AdS singleton whose flat limit reproduces the formulation in Section~\ref{sec:exotic}.
The ambient formulation
provides a natural setting for deriving it.

\subsection{Ambient description}

Our starting point is the ambient description \eqref{eq:ambient-relativistic} of the singleton in $\mathbb R^{d+1,2}$ with Cartesian coordinates $(X^a,X^{0'})=(x^a,R\,w)$,
\be \label{eq:singleton-w}
\left(\partial_a \partial^a - R^{-2}\partial_w{}^2\right) \Phi = 0 \,,\quad (x^a \partial_a + w \partial_w + \Delta)\,\Phi = 0 \,,\quad \Phi \simeq \Phi + \left(w^2 - R^{-2} x^2\right) \Psi \,,
\ee
where we rescaled the last equation by $-R^2$ to make the limit smooth.
The bulk of AdS spacetime is defined by $w(x)^2 = 1+\frac{x^2}{R^2}$, which is equivalent to 
\be \label{eq:AdS-hyperboloid}
w(x) = \sqrt{1+ \frac{x^2}{R^2}}
\ee
in the region $w >0$. By choosing the appropriate (doublet) formulation, we can define the flat $R \to \infty$ limit of the singleton directly in the bulk.

One can represent the \emph{singleton} in ambient space $\mathbb R^{d+1,2}$ by a doublet of fields
\be\label{ambientfieldexotic}
\Phi(x,w) = \upphi_-(x) + w(x)\,\upphi_+(x) \,,\qquad w>0\,,
\ee
where on $x^a$ are Cartesian coordinates on $\mathbb R^{d+1,1}$ and
\be\label{doubletformulation}
\Box_{{}_{\mathbb R^{d+1,1}}}\upphi_\pm(x)=0\,,\qquad \left(x^a \partial_a + \tfrac{d\pm 1}2 \right) \upphi_\pm(x) = 0 \,.
\ee
One can prove this statement by expanding the most general $\Phi$ a in a Taylor series satisfying \eqref{eq:singleton-w} in the coordinate $w$. The singleton reads
\be
\Phi(x,w) = \sum_{n\geqslant 0} \frac{w^n}{n!} \upphi_n(x) \,,
\ee
where homogeneity imposes, for all $n \geqslant 0$
\be
\left( x^a\partial_a + \Delta_- + n \right) \upphi_n = 0 \,.
\ee
Then, using the last relation of \eqref{eq:singleton-w} we can trade powers of $w^2$ for powers of $\frac{x^2}{R^2}$
\be
\Phi(x,w) \simeq \sum_{n \geqslant 0} \frac{1}{(2n)!} \left(\frac{x^2}{R^2}\right)^n\,\upphi_{2n}(x) + w \sum_{n \geqslant 0} \frac{1}{(2n+1)!} \left(\frac{x^2}{R^2}\right)^n\,\upphi_{2n+1}(x) \,,
\ee
while harmonicity imposes the recursion relations for all $n \geqslant 0$
\be
\Box_{\mathbb R^{d+1,1}} \upphi_n(x) = \frac{1}{R^2} \upphi_{n+2}(x) \,.
\ee
This means that there is a representative for the singleton taking the form
\be
\begin{split}
\Phi(x,w) &\simeq \sum_{n\geqslant 0} \frac{1}{(2n)!} x^{2n} \Box^n \upphi_0(x) + w \sum_{n \geqslant 0} \frac{1}{(2n+1)!} x^{2n} \Box^n \upphi_1(x) \\
&=: \upphi_-^\text{new}(x) + w\,\upphi_+^\text{new}(x) \,.
\end{split}
\ee
Remarkably, $\upphi_\pm^\text{new}(x)$ have scaling dimensions $\Delta_\pm$ as a consequence of the homogeneity of $\upphi_0$ and $\upphi_1$ and the fact that the operator $x^2\,\Box$ has homogeneity zero in $x$. Moreover, $\upphi_\pm^\text{new}(x)$ are both harmonic, which can be proven directly
\be
\Box_{{}_{\mathbb R^{d+1,1}}} \upphi_-^\text{new} = \sum_{n \geqslant 1} \frac{2n(d+2n) - 8n^2 -4n \Delta_-}{(2n)!}\, x^{2n-2} \Box^n \upphi_0 + \sum_{n \geqslant 0} \frac{1}{(2n)!}\, x^{2n} \Box^{n+1} \upphi_0 \,,
\ee
which vanishes precisely for $\Delta_- = \frac{d-1}{2}$, and similarly
\be
\Box_{{}_{\mathbb R^{d+1,1}}} \upphi_+^\text{new} = \sum_{n \geqslant 1} \frac{2n(d+2n) - 8n^2 -4n \Delta_+}{(2n+1)!} x^{2n-2} \Box^n \upphi_1 + \sum_{n \geqslant 0} \frac{1}{(2n+1)!} x^{2n} \Box^{n+1} \upphi_1 \,,
\ee
which vanishes precisely for $\Delta_+ = \frac{d+1}{2}$.

Note that the doublet of fields $\upphi_\pm(x)$ above verifies the same equations \eqref{doubletformulation} as the ones in the doublet formulation \eqref{eq:exotic-doublet} of the simpleton. 
One crucial point is that the ambient field \eqref{ambientfieldexotic} takes a slightly different form in these two cases. 
In fact, geometrically
AdS spacetime is located at the hyperboloid defined in Eq.~\eqref{eq:AdS-hyperboloid}, while Minkowski spacetime is located at the hyperplane $w = 1$. 

The flat limit is smooth in the above ambient formulation: sending $R \to \infty$ in \eqref{eq:singleton-w} gives
\begin{equation} \label{eq:ambient-exotic}
    \Box_{{}_{\mathbb R^{d+1,1}}} \Phi = 0 \,,\quad \left(x^a \partial_a + w\,\partial_w + \Delta \right) \Phi = 0 \,,\quad \Phi \simeq \Phi + w^2\,\Psi \,,
\end{equation}
with $\Delta = \frac{d-1}{2}$, as in the electric case. This provides another ambient description of the simpleton. The spacetime this ambient field $\Phi(w,x)$ lives on is now Galilean-like rather than Carrollian-like, because the well-defined object after the $R \to \infty$ limit is the co-metric $\eta^{ab} \partial_a \otimes \partial_b$, which is degenerate in the direction of the one-form ${\rm d}w$. Note that the formulation \eqref{eq:ambient-exotic} can also be seen as a Fourier transform of the previous ambient formulation \eqref{eq:ambient-electric} where the coordinates $(u,y)$ are mapped to $(w,x)$. More precisely, the equation $\partial_u{}^2\Phi=0$ in \eqref{eq:ambient-electric} is mapped to the equivalence relation $\Phi \simeq \Phi + w^2\,\Psi$ while the equivalence relation $\Phi \simeq \Phi + y^2$ in \eqref{eq:ambient-electric} is mapped to $\Box_{{}_{\mathbb R^{d+1,1}}}\Phi=0$.

The coordinate transformations strictly preserving the system \eqref{eq:ambient-exotic} coincide with Galilean isometries of this ambient spacetime. More explicitly, the vector fields that commute with the constraints $\partial^a \partial_a$, $x^a \partial_a + w \partial_w$ and $w^2$ are
\be
    J_{ab} = 2 x_{[a} \partial_{b]} \,,\quad P_a = w\,\partial_a \,.
\ee
The first generator is, again, exactly the one of Lorentz transformations, while the second looks like translations up to a $w$ factor (in fact they are Galilean boosts in ambient space). 

By taking into account the equivalence relation \eqref{eq:ambient-exotic}, there is no loss of generality in considering a representative $\Phi(x,w) = \upphi_-(x) + w\,\upphi_+(x)$ where the doublet of fields $\upphi_\pm$ verifies
\eqref{doubletformulation}. They can be grouped in a doublet as in \eqref{doublet}, which matches precisely with the choice of bulk representative given in Section~\ref{sec:exotic}. A similar formulation can be obtained for the simpleton.

\subsection{Intrinsic description}

It is useful to obtain the corresponding intrinsic formulation on AdS spacetime, which admits a smooth flat limit.

The AdS metric in `Minkowski-like coordinates' $x^a$ is the pullback of the ambient metric $\eta_{AB}$ along the hyperboloid defined in Eq.~\eqref{eq:AdS-hyperboloid}
\be\label{Minklike}
{\rm d} s_{\text{AdS}_{d+2}}^2 = -R^2\,{\rm d}w(x)^2 + \eta_{ab}{\rm d}x^a {\rm d}x^b = g_{ab}(x) {\rm d}x^a {\rm d}x^b \,,
\ee
with 
\be
g_{ab}(x) = \eta_{ab} - \frac{x_a x_b}{R^2 + x^2} \,,\quad g^{ab}(x) = \eta^{ab} + \frac{x^a x^b}{R^2} \,,\quad \text{det}(g) = - \frac{R^2}{R^2+x^2} \,,
\ee
which indeed describes a spacetime of constant negative curvature with radius $R$, i.e. $R_{abcd} = - \frac{1}{R^2} (g_{ac} g_{bd} - g_{bc} g_{ad})$. 
Note that there is a coordinate singularity at $x^2 = -R^2$, corresponding to the fact that these coordinates only cover half of AdS spacetime (the patch for which $X^{0'}>0$).
One can verify that the d'Alembert equation $\Box_{{}_{\mathbb R^{d+1,1}}} \upphi_\pm(x) = 0$ in Cartesian coordinates is equivalent to the Klein-Gordon equation on AdS spacetime in the above Minkowski-like coordinates
\be\label{eq:exotic-doubletAdS}
\left[\nabla^2_\text{AdS$_{d+2}$} - \frac{1}{R^2} \Delta_\pm(\Delta_\pm-d-1) \right] \upphi_\pm(x) = 0 \,,
\ee
where $\nabla^2_\text{AdS$_{d+2}$}$ represents the covariant Laplacian associated with the metric $g_{ab}(x)$. 
One can thus conclude that the bulk singleton can be described by a doublet of homogeneous scalar fields in AdS spacetime, admitting as a smooth flat limit the formulation of the bulk simpleton in terms of a doublet of homogeneous scalar fields in Minkowski spacetime (cf.\ Section~\ref{sec:exotic}).
This doublet formulation in AdS admits a coordinate-invariant formulation as the system formed by the mass-shell equation \eqref{eq:exotic-doubletAdS} together with the homogeneity condition $\left(\mathcal{L}_\xi + \tfrac{d\pm 1}2 \right) \upphi_\pm(x) = 0$ where the Euler vector field $\xi=x^a\partial_a$ admits a coordinate-invariant definition as follows. 
Note that the AdS metric takes the Kerr-Schild-like form $g^{ab} = \eta^{ab} + \frac{\xi^a \xi^b}{R^2}$ where $\xi$ is a conformal vector field for the background Minkowski metric $\eta_{ab}$ such that $\mathcal{L}_\xi\eta=2\eta$. 
This defines the Euler vector field $\xi$ up to a Killing vector field for the Minkowski metric $\eta$. 

Note that the two values of the scaling dimension $\Delta_\pm$ should not be interpreted as the two possible solutions of the mass-shell equation for a single scalar field of a given mass, but rather to the scaling dimensions of two distinct scalar fields with different values of their mass-squared $m_\pm^2 = \Delta_\pm(\Delta_\pm - d-1)/R^2\,$.
For the first scalar $\upphi_-$ with weight $\Delta_-$ one finds $m_-^2=\tfrac{(d-1)(d+3)}{(2R)^2}$, so there are two solutions to the mass-shell condition with scaling dimensions $\Delta = \Delta_- = \frac{d-1}{2}$ and $\Delta = \frac{d+3}{2}\,$. 
The homogeneity condition $\left( x^a\partial_a + \Delta_- \right) \varphi_- = 0$ tells us that only the one corresponding to the leading branch in an expansion close to the conformal boundary, is present. 
For the second scalar $\upphi_+$ with weight $\Delta_+$ one finds $m_+^2=-(\tfrac{d+1}{2R})^2$, so there is only one solution to the mass-shell equation with scaling dimension $\Delta = \Delta_+ = \frac{d+1}{2}\,$, which is indeed the one imposed by homogeneity $\left( x^a\partial_a + \Delta_+ \right) \upphi_+ = 0\,$.

The action of translations and Lorentz transformations on the doublet $\Phi$ reads
\be
P_a \sim
\begin{pmatrix}
0 & \frac{1}{R^2}(x^2 \partial_a + x_a) \\ \partial_a & 0
\end{pmatrix} ,\quad
J_{ab} \sim
\begin{pmatrix}
2x_{[a}\partial_{b]} & 0 \\ 0 & 2x_{[a}\partial_{b]}
\end{pmatrix} .
\ee
In particular, there is no invariant submodule, confirming that the singleton is an irreducible representation of the AdS isometry group.

%\bibliographystyle{JHEP}
%\bibliography{biblio}

\begin{thebibliography}{100}

\bibitem{Leblond}
J.-M.~L\'evy-Leblond, \emph{{Une nouvelle limite non-relativiste du groupe de
  Poincar\'e}}, {\emph{Annales de l'I.H.P. Phys. Theor.} {\bfseries 3} (1965)
  1}.

\bibitem{Gupta}
N.D.~Sen~Gupta, \emph{{On an Analogue of the Galileo Group}},
  \href{https://doi.org/10.1007/BF02740871}{\emph{Nuovo Cim.} {\bfseries 54}
  (1966) 512}.

\bibitem{Duval:2014uoa}
C.~Duval, G.W.~Gibbons, P.A.~Horvathy and P.M.~Zhang, \emph{{Carroll versus
  Newton and Galilei: two dual non-Einsteinian concepts of time}},
  \href{https://doi.org/10.1088/0264-9381/31/8/085016}{\emph{Class. Quant.
  Grav.} {\bfseries 31} (2014) 085016}
  [\href{https://arxiv.org/abs/1402.0657}{{\ttfamily 1402.0657}}].

\bibitem{Bagchi:2019xfx}
A.~Bagchi, A.~Mehra and P.~Nandi, \emph{{Field Theories with Conformal
  Carrollian Symmetry}},
  \href{https://doi.org/10.1007/JHEP05(2019)108}{\emph{JHEP} {\bfseries 05}
  (2019) 108} [\href{https://arxiv.org/abs/1901.10147}{{\ttfamily
  1901.10147}}].

\bibitem{Gupta:2020dtl}
N.~Gupta and N.V.~Suryanarayana, \emph{{Constructing Carrollian CFTs}},
  \href{https://doi.org/10.1007/JHEP03(2021)194}{\emph{JHEP} {\bfseries 03}
  (2021) 194} [\href{https://arxiv.org/abs/2001.03056}{{\ttfamily
  2001.03056}}].

\bibitem{Henneaux:2021yzg}
M.~Henneaux and P.~Salgado-Rebolledo, \emph{{Carroll contractions of
  Lorentz-invariant theories}},
  \href{https://doi.org/10.1007/JHEP11(2021)180}{\emph{JHEP} {\bfseries 11}
  (2021) 180} [\href{https://arxiv.org/abs/2109.06708}{{\ttfamily
  2109.06708}}].

\bibitem{deBoer:2021jej}
J.~de~Boer, J.~Hartong, N.A.~Obers, W.~Sybesma and S.~Vandoren, \emph{{Carroll
  Symmetry, Dark Energy and Inflation}},
  \href{https://doi.org/10.3389/fphy.2022.810405}{\emph{Front. in Phys.}
  {\bfseries 10} (2022) 810405}
  [\href{https://arxiv.org/abs/2110.02319}{{\ttfamily 2110.02319}}].

\bibitem{Rivera-Betancour:2022lkc}
D.~Rivera-Betancour and M.~Vilatte, \emph{{Revisiting the Carrollian scalar
  field}}, \href{https://doi.org/10.1103/PhysRevD.106.085004}{\emph{Phys. Rev.
  D} {\bfseries 106} (2022) 085004}
  [\href{https://arxiv.org/abs/2207.01647}{{\ttfamily 2207.01647}}].

\bibitem{Baiguera:2022lsw}
S.~Baiguera, G.~Oling, W.~Sybesma and B.T.~S\o{}gaard, \emph{{Conformal Carroll
  scalars with boosts}},
  \href{https://doi.org/10.21468/SciPostPhys.14.4.086}{\emph{SciPost Phys.}
  {\bfseries 14} (2023) 086}
  [\href{https://arxiv.org/abs/2207.03468}{{\ttfamily 2207.03468}}].

\bibitem{Banerjee:2023jpi}
K.~Banerjee, R.~Basu, B.~Krishnan, S.~Maulik, A.~Mehra and A.~Ray,
  \emph{{One-loop quantum effects in Carroll scalars}},
  \href{https://doi.org/10.1103/PhysRevD.108.085022}{\emph{Phys. Rev. D}
  {\bfseries 108} (2023) 085022}
  [\href{https://arxiv.org/abs/2307.03901}{{\ttfamily 2307.03901}}].

\bibitem{deBoer:2023fnj}
J.~de~Boer, J.~Hartong, N.A.~Obers, W.~Sybesma and S.~Vandoren, \emph{{Carroll
  stories}}, \href{https://doi.org/10.1007/JHEP09(2023)148}{\emph{JHEP}
  {\bfseries 09} (2023) 148}
  [\href{https://arxiv.org/abs/2307.06827}{{\ttfamily 2307.06827}}].

\bibitem{Koutrolikos:2023evq}
K.~Koutrolikos and M.~Najafizadeh, \emph{{Super-Carrollian and Super-Galilean
  Field Theories}},
  \href{https://doi.org/10.1103/PhysRevD.108.125014}{\emph{Phys. Rev. D}
  {\bfseries 108} (2023) 125014}
  [\href{https://arxiv.org/abs/2309.16786}{{\ttfamily 2309.16786}}].

\bibitem{Bergshoeff:2023vfd}
E.A.~Bergshoeff, A.~Campoleoni, A.~Fontanella, L.~Mele and J.~Rosseel,
  \emph{{Carroll Fermions}},
  \href{https://arxiv.org/abs/2312.00745}{{\ttfamily 2312.00745}}.

\bibitem{Bergshoeff:2014jla}
E.~Bergshoeff, J.~Gomis and G.~Longhi, \emph{{Dynamics of Carroll Particles}},
  \href{https://doi.org/10.1088/0264-9381/31/20/205009}{\emph{Class. Quant.
  Grav.} {\bfseries 31} (2014) 205009}
  [\href{https://arxiv.org/abs/1405.2264}{{\ttfamily 1405.2264}}].

\bibitem{Chen:2021xkw}
B.~Chen, R.~Liu and Y.-f.~Zheng, \emph{{On higher-dimensional Carrollian and
  Galilean conformal field theories}},
  \href{https://doi.org/10.21468/SciPostPhys.14.5.088}{\emph{SciPost Phys.}
  {\bfseries 14} (2023) 088}
  [\href{https://arxiv.org/abs/2112.10514}{{\ttfamily 2112.10514}}].

\bibitem{Bagchi:2022owq}
A.~Bagchi, D.~Grumiller and P.~Nandi, \emph{{Carrollian superconformal theories
  and super BMS}}, \href{https://doi.org/10.1007/JHEP05(2022)044}{\emph{JHEP}
  {\bfseries 05} (2022) 044}
  [\href{https://arxiv.org/abs/2202.01172}{{\ttfamily 2202.01172}}].

\bibitem{Bagchi:2022eui}
A.~Bagchi, A.~Banerjee, R.~Basu, M.~Islam and S.~Mondal, \emph{{Magic fermions:
  Carroll and flat bands}},
  \href{https://doi.org/10.1007/JHEP03(2023)227}{\emph{JHEP} {\bfseries 03}
  (2023) 227} [\href{https://arxiv.org/abs/2211.11640}{{\ttfamily
  2211.11640}}].

\bibitem{Figueroa-OFarrill:2023vbj}
J.~Figueroa-O'Farrill, A.~P\'erez and S.~Prohazka, \emph{{Carroll/fracton
  particles and their correspondence}},
  \href{https://doi.org/10.1007/JHEP06(2023)207}{\emph{JHEP} {\bfseries 06}
  (2023) 207} [\href{https://arxiv.org/abs/2305.06730}{{\ttfamily
  2305.06730}}].

\bibitem{Bagchi:2016bcd}
A.~Bagchi, R.~Basu, A.~Kakkar and A.~Mehra, \emph{{Flat Holography: Aspects of
  the dual field theory}},
  \href{https://doi.org/10.1007/JHEP12(2016)147}{\emph{JHEP} {\bfseries 12}
  (2016) 147} [\href{https://arxiv.org/abs/1609.06203}{{\ttfamily
  1609.06203}}].

\bibitem{Ciambelli:2018wre}
L.~Ciambelli, C.~Marteau, A.C.~Petkou, P.M.~Petropoulos and K.~Siampos,
  \emph{{Flat holography and Carrollian fluids}},
  \href{https://doi.org/10.1007/JHEP07(2018)165}{\emph{JHEP} {\bfseries 07}
  (2018) 165} [\href{https://arxiv.org/abs/1802.06809}{{\ttfamily
  1802.06809}}].

\bibitem{Dappiaggi:2005ci}
C.~Dappiaggi, V.~Moretti and N.~Pinamonti, \emph{{Rigorous steps towards
  holography in asymptotically flat spacetimes}},
  \href{https://doi.org/10.1142/S0129055X0600270X}{\emph{Rev. Math. Phys.}
  {\bfseries 18} (2006) 349}
  [\href{https://arxiv.org/abs/gr-qc/0506069}{{\ttfamily gr-qc/0506069}}].

\bibitem{Donnay:2022aba}
L.~Donnay, A.~Fiorucci, Y.~Herfray and R.~Ruzziconi, \emph{{Carrollian
  Perspective on Celestial Holography}},
  \href{https://doi.org/10.1103/PhysRevLett.129.071602}{\emph{Phys. Rev. Lett.}
  {\bfseries 129} (2022) 071602}
  [\href{https://arxiv.org/abs/2202.04702}{{\ttfamily 2202.04702}}].

\bibitem{Bagchi:2022emh}
A.~Bagchi, S.~Banerjee, R.~Basu and S.~Dutta, \emph{{Scattering Amplitudes:
  Celestial and Carrollian}},
  \href{https://doi.org/10.1103/PhysRevLett.128.241601}{\emph{Phys. Rev. Lett.}
  {\bfseries 128} (2022) 241601}
  [\href{https://arxiv.org/abs/2202.08438}{{\ttfamily 2202.08438}}].

\bibitem{Donnay:2022wvx}
L.~Donnay, A.~Fiorucci, Y.~Herfray and R.~Ruzziconi, \emph{{Bridging Carrollian
  and celestial holography}},
  \href{https://doi.org/10.1103/PhysRevD.107.126027}{\emph{Phys. Rev. D}
  {\bfseries 107} (2023) 126027}
  [\href{https://arxiv.org/abs/2212.12553}{{\ttfamily 2212.12553}}].

\bibitem{Salzer:2023jqv}
J.~Salzer, \emph{{An embedding space approach to Carrollian CFT correlators for
  flat space holography}},
  \href{https://doi.org/10.1007/JHEP10(2023)084}{\emph{JHEP} {\bfseries 10}
  (2023) 084} [\href{https://arxiv.org/abs/2304.08292}{{\ttfamily
  2304.08292}}].

\bibitem{Nguyen:2023vfz}
K.~Nguyen and P.~West, \emph{{Carrollian Conformal Fields and Flat
  Holography}}, \href{https://doi.org/10.3390/universe9090385}{\emph{Universe}
  {\bfseries 9} (2023) 385} [\href{https://arxiv.org/abs/2305.02884}{{\ttfamily
  2305.02884}}].

\bibitem{Campoleoni:2023fug}
A.~Campoleoni, A.~Delfante, S.~Pekar, P.M.~Petropoulos, D.~Rivera-Betancour and
  M.~Vilatte, \emph{{Flat from anti de Sitter}},
  \href{https://doi.org/10.1007/JHEP12(2023)078}{\emph{JHEP} {\bfseries 12}
  (2023) 078} [\href{https://arxiv.org/abs/2309.15182}{{\ttfamily
  2309.15182}}].

\bibitem{Bagchi:2023cen}
A.~Bagchi, P.~Dhivakar and S.~Dutta, \emph{{Holography in Flat Spacetimes: the
  case for Carroll}},  \href{https://arxiv.org/abs/2311.11246}{{\ttfamily
  2311.11246}}.

\bibitem{Mason:2023mti}
L.~Mason, R.~Ruzziconi and A.~Yelleshpur~Srikant, \emph{{Carrollian amplitudes
  and celestial symmetries}},
  \href{https://doi.org/10.1007/JHEP05(2024)012}{\emph{JHEP} {\bfseries 05}
  (2024) 012} [\href{https://arxiv.org/abs/2312.10138}{{\ttfamily
  2312.10138}}].

\bibitem{Pasterski:2021raf}
S.~Pasterski, M.~Pate and A.-M.~Raclariu, \emph{{Celestial Holography}},  in
  \emph{{2022 Snowmass Summer Study}}, 11, 2021
  [\href{https://arxiv.org/abs/2111.11392}{{\ttfamily 2111.11392}}].

\bibitem{Donnay:2023mrd}
L.~Donnay, \emph{{Celestial holography: An asymptotic symmetry perspective}},
  \href{https://doi.org/10.1016/j.physrep.2024.04.003}{\emph{Phys. Rept.}
  {\bfseries 1073} (2024) 1}
  [\href{https://arxiv.org/abs/2310.12922}{{\ttfamily 2310.12922}}].

\bibitem{Sundborg:2000wp}
B.~Sundborg, \emph{{Stringy gravity, interacting tensionless strings and
  massless higher spins}},
  \href{https://doi.org/10.1016/S0920-5632(01)01545-6}{\emph{Nucl. Phys. B
  Proc. Suppl.} {\bfseries 102} (2001) 113}
  [\href{https://arxiv.org/abs/hep-th/0103247}{{\ttfamily hep-th/0103247}}].

\bibitem{Witten_talk}
E.~Witten, ``Spacetime reconstruction.'' Talk given at the \emph{John Schwarz
  60th birthday symposium}, Pasadena (U.S.A), 2001.

\bibitem{Mikhailov:2002bp}
A.~Mikhailov, \emph{{Notes on higher spin symmetries}},
  \href{https://arxiv.org/abs/hep-th/0201019}{{\ttfamily hep-th/0201019}}.

\bibitem{Sezgin:2002rt}
E.~Sezgin and P.~Sundell, \emph{{Massless higher spins and holography}},
  \href{https://doi.org/10.1016/S0550-3213(02)00739-3}{\emph{Nucl. Phys. B}
  {\bfseries 644} (2002) 303}
  [\href{https://arxiv.org/abs/hep-th/0205131}{{\ttfamily hep-th/0205131}}].

\bibitem{Klebanov:2002ja}
I.R.~Klebanov and A.M.~Polyakov, \emph{{AdS dual of the critical O(N) vector
  model}}, \href{https://doi.org/10.1016/S0370-2693(02)02980-5}{\emph{Phys.
  Lett. B} {\bfseries 550} (2002) 213}
  [\href{https://arxiv.org/abs/hep-th/0210114}{{\ttfamily hep-th/0210114}}].

\bibitem{Giombi:2016ejx}
S.~Giombi, \emph{{Higher Spin \textemdash{} CFT Duality}},  in
  \emph{{Theoretical Advanced Study Institute in Elementary Particle Physics}:
  {New Frontiers in Fields and Strings}}, pp.~137--214, 2017,
  \href{https://doi.org/10.1142/9789813149441_0003}{DOI}
  [\href{https://arxiv.org/abs/1607.02967}{{\ttfamily 1607.02967}}].

\bibitem{Bekaert:2010hw}
X.~Bekaert, N.~Boulanger and P.~Sundell, \emph{{How higher-spin gravity
  surpasses the spin two barrier: no-go theorems versus yes-go examples}},
  \href{https://doi.org/10.1103/RevModPhys.84.987}{\emph{Rev. Mod. Phys.}
  {\bfseries 84} (2012) 987} [\href{https://arxiv.org/abs/1007.0435}{{\ttfamily
  1007.0435}}].

\bibitem{Ponomarev:2016lrm}
D.~Ponomarev and E.D.~Skvortsov, \emph{{Light-Front Higher-Spin Theories in
  Flat Space}}, \href{https://doi.org/10.1088/1751-8121/aa56e7}{\emph{J. Phys.
  A} {\bfseries 50} (2017) 095401}
  [\href{https://arxiv.org/abs/1609.04655}{{\ttfamily 1609.04655}}].

\bibitem{Krasnov:2021nsq}
K.~Krasnov, E.~Skvortsov and T.~Tran, \emph{{Actions for self-dual Higher Spin
  Gravities}}, \href{https://doi.org/10.1007/JHEP08(2021)076}{\emph{JHEP}
  {\bfseries 08} (2021) 076}
  [\href{https://arxiv.org/abs/2105.12782}{{\ttfamily 2105.12782}}].

\bibitem{Ren:2022sws}
L.~Ren, M.~Spradlin, A.~Yelleshpur~Srikant and A.~Volovich, \emph{{On effective
  field theories with celestial duals}},
  \href{https://doi.org/10.1007/JHEP08(2022)251}{\emph{JHEP} {\bfseries 08}
  (2022) 251} [\href{https://arxiv.org/abs/2206.08322}{{\ttfamily
  2206.08322}}].

\bibitem{Ponomarev:2022ryp}
D.~Ponomarev, \emph{{Towards higher-spin holography in flat space}},
  \href{https://doi.org/10.1007/JHEP01(2023)084}{\emph{JHEP} {\bfseries 01}
  (2023) 084} [\href{https://arxiv.org/abs/2210.04035}{{\ttfamily
  2210.04035}}].

\bibitem{Ponomarev:2022qkx}
D.~Ponomarev, \emph{{Chiral higher-spin holography in flat space: the
  Flato-Fronsdal theorem and lower-point functions}},
  \href{https://doi.org/10.1007/JHEP01(2023)048}{\emph{JHEP} {\bfseries 01}
  (2023) 048} [\href{https://arxiv.org/abs/2210.04036}{{\ttfamily
  2210.04036}}].

\bibitem{Monteiro:2022xwq}
R.~Monteiro, \emph{{From Moyal deformations to chiral higher-spin theories and
  to celestial algebras}},
  \href{https://doi.org/10.1007/JHEP03(2023)062}{\emph{JHEP} {\bfseries 03}
  (2023) 062} [\href{https://arxiv.org/abs/2212.11266}{{\ttfamily
  2212.11266}}].

\bibitem{Campoleoni:2021blr}
A.~Campoleoni and S.~Pekar, \emph{{Carrollian and Galilean conformal
  higher-spin algebras in any dimensions}},
  \href{https://doi.org/10.1007/JHEP02(2022)150}{\emph{JHEP} {\bfseries 02}
  (2022) 150} [\href{https://arxiv.org/abs/2110.07794}{{\ttfamily
  2110.07794}}].

\bibitem{Boulanger:2023prx}
N.~Boulanger, A.~Campoleoni and S.~Pekar, \emph{{New higher-spin curvatures in
  flat space}}, \href{https://doi.org/10.1103/PhysRevD.108.L101904}{\emph{Phys.
  Rev. D} {\bfseries 108} (2023) L101904}
  [\href{https://arxiv.org/abs/2306.05367}{{\ttfamily 2306.05367}}].

\bibitem{Bekaert:2022oeh}
X.~Bekaert, A.~Campoleoni and S.~Pekar, \emph{{Carrollian conformal scalar as
  flat-space singleton}},
  \href{https://doi.org/10.1016/j.physletb.2023.137734}{\emph{Phys. Lett. B}
  {\bfseries 838} (2023) 137734}
  [\href{https://arxiv.org/abs/2211.16498}{{\ttfamily 2211.16498}}].

\bibitem{Dirac:1963ta}
P.A.M.~Dirac, \emph{{A Remarkable representation of the 3 + 2 de Sitter
  group}}, \href{https://doi.org/10.1063/1.1704016}{\emph{J. Math. Phys.}
  {\bfseries 4} (1963) 901}.

\bibitem{Fronsdal:1978vb}
C.~Fronsdal, \emph{{Singletons and Massless, Integral Spin Fields on de Sitter
  Space (Elementary Particles in a Curved Space. 7.}},
  \href{https://doi.org/10.1103/PhysRevD.20.848}{\emph{Phys. Rev. D} {\bfseries
  20} (1979) 848}.

\bibitem{Angelopoulos:1980wg}
E.~Angelopoulos, M.~Flato, C.~Fronsdal and D.~Sternheimer, \emph{{Massless
  Particles, Conformal Group and De Sitter Universe}},
  \href{https://doi.org/10.1103/PhysRevD.23.1278}{\emph{Phys. Rev. D}
  {\bfseries 23} (1981) 1278}.

\bibitem{Flato:1978qz}
M.~Flato and C.~Fronsdal, \emph{{One Massless Particle Equals Two Dirac
  Singletons: Elementary Particles in a Curved Space. 6.}},
  \href{https://doi.org/10.1007/BF00400170}{\emph{Lett. Math. Phys.} {\bfseries
  2} (1978) 421}.

\bibitem{Ponomarev:2021xdq}
D.~Ponomarev, \emph{{3d conformal fields with manifest sl(2,
  \ensuremath{\mathbb{C}})}},
  \href{https://doi.org/10.1007/JHEP06(2021)055}{\emph{JHEP} {\bfseries 06}
  (2021) 055} [\href{https://arxiv.org/abs/2104.02770}{{\ttfamily
  2104.02770}}].

\bibitem{Henneaux:1979vn}
M.~Henneaux, \emph{{Geometry of Zero Signature Space-times}}, {\emph{Bull. Soc.
  Math. Belg.} {\bfseries 31} (1979) 47}.

\bibitem{Campiglia:2014yka}
M.~Campiglia and A.~Laddha, \emph{{Asymptotic symmetries and subleading soft
  graviton theorem}},
  \href{https://doi.org/10.1103/PhysRevD.90.124028}{\emph{Phys. Rev. D}
  {\bfseries 90} (2014) 124028}
  [\href{https://arxiv.org/abs/1408.2228}{{\ttfamily 1408.2228}}].

\bibitem{Duval:2014uva}
C.~Duval, G.W.~Gibbons and P.A.~Horvathy, \emph{{Conformal Carroll groups and
  BMS symmetry}},
  \href{https://doi.org/10.1088/0264-9381/31/9/092001}{\emph{Class. Quant.
  Grav.} {\bfseries 31} (2014) 092001}
  [\href{https://arxiv.org/abs/1402.5894}{{\ttfamily 1402.5894}}].

\bibitem{Wigner:1939cj}
E.P.~Wigner, \emph{{On Unitary Representations of the Inhomogeneous Lorentz
  Group}}, \href{https://doi.org/10.2307/1968551}{\emph{Annals Math.}
  {\bfseries 40} (1939) 149}.

\bibitem{McCarthy01}
P.J.~McCarthy, \emph{{Representations of the Bondi-Metzner-Sachs Group. I.
  Determination of the Representations}},
  \href{https://doi.org/10.1098/rspa.1972.0157}{\emph{Proc. R. Soc. Lond. A}
  {\bfseries 330} (1972) 517}.

\bibitem{McCarthy317}
P.J.~McCarthy, \emph{{Representations of the Bondi-Metzner-Sachs Group. II.
  Properties and Classification of the Representations}},
  \href{https://doi.org/10.1098/rspa.1973.0065}{\emph{Proc. R. Soc. Lond. A}
  {\bfseries 333} (1973) 317}.

\bibitem{Sachs:1962wk}
R.K.~Sachs, \emph{{Gravitational waves in general relativity. 8. Waves in
  asymptotically flat space-times}},
  \href{https://doi.org/10.1098/rspa.1962.0206}{\emph{Proc. R. Soc. Lond. A}
  {\bfseries 270} (1962) 103}.

\bibitem{Bekaert:2022ipg}
X.~Bekaert and B.~Oblak, \emph{{Massless scalars and higher-spin BMS in any
  dimension}}, \href{https://doi.org/10.1007/JHEP11(2022)022}{\emph{JHEP}
  {\bfseries 11} (2022) 022}
  [\href{https://arxiv.org/abs/2209.02253}{{\ttfamily 2209.02253}}].

\bibitem{Bacry:1968zf}
H.~Bacry and J.~Levy-Leblond, \emph{{Possible kinematics}},
  \href{https://doi.org/10.1063/1.1664490}{\emph{J. Math. Phys.} {\bfseries 9}
  (1968) 1605}.

\bibitem{Figueroa-OFarrill:2018ilb}
J.~Figueroa-O'Farrill and S.~Prohazka, \emph{{Spatially isotropic homogeneous
  spacetimes}}, \href{https://doi.org/10.1007/JHEP01(2019)229}{\emph{JHEP}
  {\bfseries 01} (2019) 229}
  [\href{https://arxiv.org/abs/1809.01224}{{\ttfamily 1809.01224}}].

\bibitem{Herfray:2021qmp}
Y.~Herfray, \emph{{Carrollian manifolds and null infinity: a view from Cartan
  geometry}}, \href{https://doi.org/10.1088/1361-6382/ac635f}{\emph{Class.
  Quant. Grav.} {\bfseries 39} (2022) 215005}
  [\href{https://arxiv.org/abs/2112.09048}{{\ttfamily 2112.09048}}].

\bibitem{Satishchandran:2019pyc}
G.~Satishchandran and R.M.~Wald, \emph{{Asymptotic behavior of massless fields
  and the memory effect}},
  \href{https://doi.org/10.1103/PhysRevD.99.084007}{\emph{Phys. Rev. D}
  {\bfseries 99} (2019) 084007}
  [\href{https://arxiv.org/abs/1901.05942}{{\ttfamily 1901.05942}}].

\bibitem{Campoleoni:2020ejn}
A.~Campoleoni, D.~Francia and C.~Heissenberg, \emph{{On asymptotic symmetries
  in higher dimensions for any spin}},
  \href{https://doi.org/10.1007/JHEP12(2020)129}{\emph{JHEP} {\bfseries 12}
  (2020) 129} [\href{https://arxiv.org/abs/2011.04420}{{\ttfamily
  2011.04420}}].

\bibitem{Figueroa-OFarrill:2023qty}
J.~Figueroa-O'Farrill, A.~P\'erez and S.~Prohazka, \emph{{Quantum
  Carroll/fracton particles}},
  \href{https://doi.org/10.1007/JHEP10(2023)041}{\emph{JHEP} {\bfseries 10}
  (2023) 041} [\href{https://arxiv.org/abs/2307.05674}{{\ttfamily
  2307.05674}}].

\bibitem{Campoleoni:2023eqp}
A.~Campoleoni, A.~Delfante, D.~Francia and C.~Heissenberg,
  \emph{{Renormalization of spin-one asymptotic charges in AdS$_{D}$}},
  \href{https://doi.org/10.1007/JHEP12(2023)061}{\emph{JHEP} {\bfseries 12}
  (2023) 061} [\href{https://arxiv.org/abs/2308.00476}{{\ttfamily
  2308.00476}}].

\bibitem{Bekaert:2012ux}
X.~Bekaert, E.~Joung and J.~Mourad, \emph{{Comments on higher-spin
  holography}}, \href{https://doi.org/10.1002/prop.201200014}{\emph{Fortsch.
  Phys.} {\bfseries 60} (2012) 882}
  [\href{https://arxiv.org/abs/1202.0543}{{\ttfamily 1202.0543}}].

\bibitem{Bekaert:2012vt}
X.~Bekaert and M.~Grigoriev, \emph{{Notes on the ambient approach to boundary
  values of AdS gauge fields}},
  \href{https://doi.org/10.1088/1751-8113/46/21/214008}{\emph{J. Phys. A}
  {\bfseries 46} (2013) 214008}
  [\href{https://arxiv.org/abs/1207.3439}{{\ttfamily 1207.3439}}].

\bibitem{Leigh:2012mz}
R.G.~Leigh and A.C.~Petkou, \emph{{Singleton deformation of higher-spin theory
  and the phase structure of the three-dimensional O(N) vector model}},
  \href{https://doi.org/10.1103/PhysRevD.88.046006}{\emph{Phys. Rev. D}
  {\bfseries 88} (2013) 046006}
  [\href{https://arxiv.org/abs/1212.4421}{{\ttfamily 1212.4421}}].

\bibitem{Bekaert:2013zya}
X.~Bekaert and M.~Grigoriev, \emph{{Higher order singletons, partially massless
  fields and their boundary values in the ambient approach}},
  \href{https://doi.org/10.1016/j.nuclphysb.2013.08.015}{\emph{Nucl. Phys. B}
  {\bfseries 876} (2013) 667}
  [\href{https://arxiv.org/abs/1305.0162}{{\ttfamily 1305.0162}}].

\bibitem{Kobayashi_2003}
T.~Kobayashi and B.~Ørsted, \emph{Analysis on the minimal representation of
  o(p,q) i. realization via conformal geometry},
  \href{https://doi.org/10.1016/s0001-8708(03)00012-4}{\emph{Advances in
  Mathematics} {\bfseries 180} (2003) 486–512}.

\bibitem{Basile:2017kaz}
T.~Basile, \emph{{A note on rectangular partially massless fields}},
  \href{https://doi.org/10.3390/universe4010004}{\emph{Universe} {\bfseries 4}
  (2018) 4} [\href{https://arxiv.org/abs/1710.10572}{{\ttfamily 1710.10572}}].

\bibitem{Bagchi:2012yk}
A.~Bagchi, S.~Detournay and D.~Grumiller, \emph{{Flat-Space Chiral Gravity}},
  \href{https://doi.org/10.1103/PhysRevLett.109.151301}{\emph{Phys. Rev. Lett.}
  {\bfseries 109} (2012) 151301}
  [\href{https://arxiv.org/abs/1208.1658}{{\ttfamily 1208.1658}}].

\bibitem{Fiorucci:2023lpb}
A.~Fiorucci, D.~Grumiller and R.~Ruzziconi, \emph{{Logarithmic celestial
  conformal field theory}},
  \href{https://doi.org/10.1103/PhysRevD.109.L021902}{\emph{Phys. Rev. D}
  {\bfseries 109} (2024) L021902}
  [\href{https://arxiv.org/abs/2305.08913}{{\ttfamily 2305.08913}}].

\bibitem{Figueroa-OFarrill:2021sxz}
J.~Figueroa-O'Farrill, E.~Have, S.~Prohazka and J.~Salzer, \emph{{Carrollian
  and celestial spaces at infinity}},
  \href{https://doi.org/10.1007/JHEP09(2022)007}{\emph{JHEP} {\bfseries 09}
  (2022) 007} [\href{https://arxiv.org/abs/2112.03319}{{\ttfamily
  2112.03319}}].

\bibitem{Have:2024dff}
E.~Have, K.~Nguyen, S.~Prohazka and J.~Salzer, \emph{{Massive carrollian fields
  at timelike infinity}},  \href{https://arxiv.org/abs/2402.05190}{{\ttfamily
  2402.05190}}.

\bibitem{Barnich:2012rz}
G.~Barnich, A.~Gomberoff and H.A.~Gonz\'alez, \emph{{Three-dimensional
  Bondi-Metzner-Sachs invariant two-dimensional field theories as the flat
  limit of Liouville theory}},
  \href{https://doi.org/10.1103/PhysRevD.87.124032}{\emph{Phys. Rev. D}
  {\bfseries 87} (2013) 124032}
  [\href{https://arxiv.org/abs/1210.0731}{{\ttfamily 1210.0731}}].

\bibitem{Barnich:2014xnb}
G.~Barnich, A.~Gomberoff and H.A.~Gonz\'alez, \emph{{A 2D field theory
  equivalent to 3D gravity with no cosmological constant}},
  \href{https://doi.org/10.1007/978-3-642-40157-2_11}{\emph{Springer Proc.
  Math. Stat.} {\bfseries 60} (2014) 135}
  [\href{https://arxiv.org/abs/1303.3568}{{\ttfamily 1303.3568}}].

\bibitem{Giombi:2013yva}
S.~Giombi, I.R.~Klebanov, S.S.~Pufu, B.R.~Safdi and G.~Tarnopolsky, \emph{{AdS
  Description of Induced Higher-Spin Gauge Theory}},
  \href{https://doi.org/10.1007/JHEP10(2013)016}{\emph{JHEP} {\bfseries 10}
  (2013) 016} [\href{https://arxiv.org/abs/1306.5242}{{\ttfamily 1306.5242}}].

\bibitem{Pretko:2018jbi}
M.~Pretko, \emph{{The Fracton Gauge Principle}},
  \href{https://doi.org/10.1103/PhysRevB.98.115134}{\emph{Phys. Rev. B}
  {\bfseries 98} (2018) 115134}
  [\href{https://arxiv.org/abs/1807.11479}{{\ttfamily 1807.11479}}].

\bibitem{Bidussi:2021nmp}
L.~Bidussi, J.~Hartong, E.~Have, J.~Musaeus and S.~Prohazka, \emph{{Fractons,
  dipole symmetries and curved spacetime}},
  \href{https://doi.org/10.21468/SciPostPhys.12.6.205}{\emph{SciPost Phys.}
  {\bfseries 12} (2022) 205}
  [\href{https://arxiv.org/abs/2111.03668}{{\ttfamily 2111.03668}}].

\bibitem{Bekaert:2011js}
X.~Bekaert, \emph{{Singletons and their maximal symmetry algebras}},  in
  \emph{{6th Summer School in Modern Mathematical Physics}}, pp.~71--89, 11,
  2011 [\href{https://arxiv.org/abs/1111.4554}{{\ttfamily 1111.4554}}].

\bibitem{Deser:1976iy}
S.~Deser and C.~Teitelboim, \emph{{Duality Transformations of Abelian and
  Nonabelian Gauge Fields}},
  \href{https://doi.org/10.1103/PhysRevD.13.1592}{\emph{Phys. Rev. D}
  {\bfseries 13} (1976) 1592}.

\bibitem{Henneaux:2004jw}
M.~Henneaux and C.~Teitelboim, \emph{{Duality in linearized gravity}},
  \href{https://doi.org/10.1103/PhysRevD.71.024018}{\emph{Phys. Rev. D}
  {\bfseries 71} (2005) 024018}
  [\href{https://arxiv.org/abs/gr-qc/0408101}{{\ttfamily gr-qc/0408101}}].

\bibitem{Henneaux:2016zlu}
M.~Henneaux, S.~H\"ortner and A.~Leonard, \emph{{Twisted self-duality for
  higher spin gauge fields and prepotentials}},
  \href{https://doi.org/10.1103/PhysRevD.94.105027}{\emph{Phys. Rev. D}
  {\bfseries 94} (2016) 105027}
  [\href{https://arxiv.org/abs/1609.04461}{{\ttfamily 1609.04461}}].

\bibitem{Duval:2014lpa}
C.~Duval, G.W.~Gibbons and P.A.~Horvathy, \emph{{Conformal Carroll groups}},
  \href{https://doi.org/10.1088/1751-8113/47/33/335204}{\emph{J. Phys. A}
  {\bfseries 47} (2014) 335204}
  [\href{https://arxiv.org/abs/1403.4213}{{\ttfamily 1403.4213}}].

\end{thebibliography}

\end{document}